\documentclass[twocolumn,traditabstract]{aa}  
\usepackage{fixltx2e}
\usepackage[english]{babel}
\usepackage{graphicx,amsmath}
\usepackage{epstopdf}
\usepackage{epsf,color}
\usepackage[mathscr]{eucal}
\usepackage{amsmath}
\usepackage{amssymb,amsfonts}
\usepackage{natbib}
\usepackage{graphicx}
\usepackage{txfonts}
\usepackage{dsfont}
\definecolor{Mygreen}{rgb}{0.00, 0.5, 0.5}
\definecolor{Mypink}{rgb}{1.0, 0.0, 0.5}
\definecolor{Myblue}{rgb}{0.00, 0.2, 0.8}
\definecolor{Myred}{rgb}{0.80, 0.2, 0.0}
\usepackage[breaklinks, citecolor=Myblue, linkcolor=Mygreen, urlcolor=Mygreen, colorlinks=true, debug, baseurl=' ']{hyperref}
\usepackage{float}
\usepackage{color}

\def\simlt{\lower.5ex\hbox{$\; \buildrel < \over \sim \;$}}
\def\simgt{\lower.5ex\hbox{$\; \buildrel > \over \sim \;$}}

\bibpunct{(}{)}{ and}{a}{}{,}
\newfont{\gwpfont}{cmssq8 scaled 1000}

\begin{document}

\def\aj{AJ}%
\def\araa{ARA\&A}%
\def\apj{ApJ}%
\def\apjl{ApJ}%
\def\apjs{ApJS}%
\def\aap{A\&A}%
 \def\aapr{A\&A~Rev.}%
\def\aaps{A\&AS}%
\def\mnras{MNRAS}
\def\ssr{SSRv}
\def\nat{Nature}
\def\jcap{JCAP}

\def\Mgv{M_{\rm g,500}}
\def\Mg{M_{\rm g}}
\def\YX {Y_{\rm X}}
\def\LXv {L_{\rm X,500}}
\def\TX {T_{\rm X}}
\def\fgv {f_{\rm g,500}}
\def\fg  {f_{\rm g}}
\def\kT {{\rm k}T}
\def\ne {n_{\rm e}}
\def\Mv {M_{\rm 500}}
\def \Rv {R_{500}}
\def\keV {\rm keV}
\def\Yv{Y_{500}}

\def\MT {$M$--$T_{\rm X}$}
\def\MYX {$M$--$Y_{\rm X}$}
\def\MMg {$M_{500}$--$M_{\rm g,500}$}
\def\MgT {$M_{\rm g,500}$--$T_{\rm X}$}
\def\MgY {$M_{\rm g,500}$--$Y_{\rm X}$}

\def\msol {{\rm M_{\odot}}}

\def\lesssim{\mathrel{\hbox{\rlap{\hbox{\lower4pt\hbox{$\sim$}}}\hbox{$<$}}}}
\def\gtrsim{\mathrel{\hbox{\rlap{\hbox{\lower4pt\hbox{$\sim$}}}\hbox{$>$}}}}

\def\psz{PSZ2\,G144.83$+$25.11}

\def\xmm{XMM-{\it Newton}}
\def\planck{{\it Planck}} 
\def\chandra{{\it Chandra}}
\def \rosat {\hbox{\it ROSAT}}
\newcommand{\excpres}{{\gwpfont EXCPRES}}
\newcommand{\ma}[1]{\textcolor{red}{{ #1}}}
\title{Impact of ICM disturbances on the mean pressure profile of galaxy clusters: A prospective study of the NIKA2 SZ large program with MUSIC synthetic clusters}

\author{version 1.0}

\author{F.~Ruppin \inst{\ref{LPSC},\ref{MIT}}
\and  F.~Sembolini \inst{\ref{Roma},\ref{Madrid},\ref{Madrid2}}
\and  M.~De~Petris \inst{\ref{Roma}}
\and  R.~Adam \inst{\ref{LLR}}
\and  G.~Cialone \inst{\ref{Roma}}
\and  J.F.~Mac\'ias-P\'erez \inst{\ref{LPSC}}
\and  F.~Mayet \inst{\ref{LPSC}}
\and  L.~Perotto \inst{\ref{LPSC}}
\and  G.~Yepes \inst{\ref{Madrid},\ref{Madrid2}}}

\institute{
Univ. Grenoble Alpes, CNRS, LPSC/IN2P3, 53 avenue des Martyrs, 38000 Grenoble, France
  \label{LPSC}
\and
Kavli Institute for Astrophysics and Space Research, Massachusetts Institute of Technology, Cambridge, MA 02139, USA
 \label{MIT}
\and
Dipartimento di Fisica, Sapienza Universit\`a di Roma, Piazzale Aldo Moro 5, I-00185 Roma, Italy
  \label{Roma}
\and
Departamento de F\'isica Te\'orica, M\'odulo 8, Facultad de Ciencias, Universidad Aut\'onoma de Madrid, E-28049 Cantoblanco, Madrid, Spain
\label{Madrid}
\and
Centro de Investigaci\'on Avanzada en F\'{\i}sica Fundamental (CIAFF), Universidad Aut\'onoma de Madrid, E-28049 Madrid, Spain
\label{Madrid2}
\and
Laboratoire Leprince-Ringuet, \'Ecole Polytechnique, CNRS/IN2P3, 91128 Palaiseau, France
\label{LLR}
}

\abstract {\emph{Context}. The mean pressure profile of the galaxy cluster population plays an essential role in cosmological analyses. An accurate characterization of the shape, intrinsic scatter, and redshift evolution of this profile is necessary to estimate some of the biases and systematic effects that currently prevent cosmological analyses based on thermal Sunyaev-Zel'dovich (tSZ) surveys from obtaining precise and unbiased cosmological constraints. This is one of the main goals of the ongoing NIKA2 SZ large program, which aims at mapping the tSZ signal of a representative cluster sample selected from the \planck\ and ACT catalogs and spans a redshift range $0.5 < z < 0.9$.\\
\emph{Aims}. To estimate the impact of intracluster medium (ICM) disturbances that can be detected by NIKA2 on the mean pressure profile of galaxy clusters, we realized a study based on a synthetic cluster sample that is similar to that of the NIKA2 SZ large program.\\
\emph{Methods}. To reach this goal we employed the hydrodynamical N-body simulation \emph{Marenostrum MUltidark SImulations of galaxy Clusters} (MUSIC). We simulated realistic NIKA2 and \planck\ tSZ observations, which were jointly analyzed to estimate the ICM pressure profile of each cluster. A comparison of the deprojected profiles with the true radial profiles directly extracted from the MUSIC simulation allowed us to validate the NIKA2 tSZ pipeline and to study the impact of ICM disturbances on the characterization of the ICM pressure distribution even at high redshift. After normalizing each profile by the integrated quantities estimated under the hydrostatic equilibrium hypothesis, we evaluated the mean pressure profile of the twin sample and show that it is compatible with that extracted directly from the MUSIC simulation in the scale range that can be recovered by NIKA2. We studied the impact of cluster dynamical state on both its shape and associated scatter.\\
\emph{Results}. We observe that at $\mathrm{R_{500}}$ the scatter of the distribution of normalized pressure profiles associated with the selected morphologically disturbed clusters is 65\% larger than that associated with relaxed clusters. Furthermore, we show that using a basic modeling of the thermal pressure distribution in the deprojection procedure induces a significant increase of the scatter associated with the mean normalized pressure profile compared to the true distribution extracted directly from the simulation.\\ 
\emph{Conclusions}. We conclude that the NIKA2 SZ large program will facilitate characterization of the potential redshift evolution of the mean pressure profile properties due to the performance of the NIKA2 camera, thereby allowing for a precise measurement of cluster morphology and ICM thermodynamic properties up to $\mathrm{R_{500}}$ at high redshift.}

\titlerunning{Simulation of the NIKA2 SZ large program with MUSIC clusters}
\authorrunning{F. Ruppin \emph{et al.}}
\maketitle

\section{Introduction}\label{sec:Introduction}

In the concordance model of cosmology, the hierarchical growth of structures culminates with the formation of galaxy clusters. These gravitationally bound objects are the most recent to form and can therefore be used as powerful probes of the intrinsic properties of the universe during the latest stage of its evolution  \citep[{e.g.,}][]{voi05}. Galaxy clusters are thus highly complementary to geometrical probes such as the primary anisotropies in temperature and polarization of the cosmic microwave background \citep[CMB;][]{pla18}, baryon acoustic oscillations \citep{and14}, or type Ia supernovae \citep{per97}, which all enable exploring different epochs of the universe history. For example, the abundance of galaxy clusters and its evolution with resdshift provides constraints on the total matter density of the universe $\Omega_m$ and the amplitude of the linear matter power spectrum at a scale of $8 h^{-1}$~Mpc, $\sigma_8$, \citep[{e.g.,}][]{pla16a}. A comparison of the cosmological constraints established from the study of the statistical properties of galaxy cluster with those obtained using high redshift probes provides valuable information on the large scale structure formation processes and may highlight potential defects of the standard cosmological model \citep[{e.g.,} ][]{boe16}.\\
\indent Although most of the matter content of galaxy clusters is under the form of dark matter, about 12\% of the total mass of these clusters is constituted of hot gas trapped in their gravitational potential well called the intracluster medium (ICM). This gas leaves an imprint on the CMB known as the Sunyaev-Zel'dovich effect \citep[SZ;][]{sun72}, which is a redshift independent probe that can be used to detect galaxy clusters and study their ICM properties up to high redshift. The thermal SZ effect (tSZ) can be used to measure directly the ICM pressure distribution. Furthermore, the integrated tSZ signal over a solid angle is proportional to the thermal energy content of galaxy clusters, which is expected to be closely related to the overall cluster mass \citep[see][]{arn10,pla14}. The tSZ effect is therefore an excellent cosmological probe that enables the establishment of nearly mass-limited samples of galaxy clusters on a wide redshift range \citep[e.g.,][]{pla16b} and that provides a low-scatter mass proxy that can be used to constrain the mass function and its evolution with redshift.\\
\indent At present, the systematic biases and uncertainties affecting the measurement of the cluster thermodynamic properties represent the main limiting factors of cosmological constraints derived from the study of the statistical properties of the cluster population \citep[{e.g.,}][]{sal18}. These systematic uncertainties are caused by projection effects due to unidentified substructures along the line of sight, by the presence of nonthermal pressure support within unvirialized regions of the ICM ({e.g.}, supernovae induced winds, gas turbulence, and feedback from active galactic nuclei) or by deviations from hydrostatic equilibrium ({e.g.,} shocks due to merging events; see \citealt[][for a review]{mro18}). The mean pressure profile of the cluster population plays a fundamental role among the different components of a cosmological analysis based on a tSZ survey. The overall amplitude and shape of the mean pressure profile define those of the matched filter used to define a catalog of galaxy clusters and estimate their integrated tSZ signal \citep[{e.g.,}][]{mel06}. The mean pressure profile properties also characterize the global amplitude of the tSZ power spectrum and its shape at high multipole \citep[{e.g.,}][]{bol18}. At redshifts $0.5 < z < 1$, the normalized radial range $R/R_{500} \in [0.1,1]$, where the transition between the inner and outer slopes of the pressure profile of a typical cluster of mass $M_{500} = 5\times 10^{14}~\mathrm{M_{\odot}}$ is expected, corresponds to angular scales between 10~arcsec and 2.5~arcmin, respectively. Mapping the tSZ signal of galaxy clusters at an angular resolution better than an arcminute is therefore a mandatory step in performing an accurate characterization of the mean pressure profile properties, which is essential to realize unbiased cosmological analyses using results from tSZ surveys. Indeed, a detailed study of the spatial distribution of the tSZ signal within clusters will enable us to improve our knowledge of the mean pressure profile, and help reduce the systematic uncertainties and part of the biases associated with the current mass estimates based on this probe.\\
\indent The new IRAM KIDs Array 2 (NIKA2) camera installed at the Institut de Radio Astronomie Millimetrique (IRAM) 30 m telescope on Pico Veleta (Spain) and its pathfinder NIKA have already proven to be excellent instruments to map the tSZ signal of intermediate and high redshift galaxy clusters (see \citealt{ada14,ada15,ada16a,ada17a,ada17b,ada18b}, \citealt{rup17}, and \citealt{rom17}). In particular, the first tSZ observations with NIKA2 \citep{rup18} have shown that the characteristics of the camera in terms of angular resolution and field-of-view extension make it possible to identify disturbed regions within galaxy clusters and study their impact on the ICM pressure distribution.\\
\indent The NIKA2 SZ large program consists in mapping the tSZ signal of a representative sample of 50 galaxy clusters at high angular resolution and in the $0.5 < z < 0.9$ redshift range \citep{com16}. The main goal of this project is to constrain the mean pressure profile and the scaling relation that links the tSZ observable to the cluster mass. The tSZ data measured by NIKA2 will be used jointly with X-ray observations made by the \xmm\ observatory on the same sample in order to constrain the hydrostatic mass of each cluster. In addition, these multi-probe analyses will allow us to study all the ICM thermodynamic properties and thus to understand the phenomenons that lead to the physical evolution of massive halos in the universe through the accretion processes and merger events with subclusters. This will result in a better characterization of the tSZ-mass scaling relation, mean pressure profile, and their potential evolution with redshift. Furthermore, the NIKA2 SZ large program will enable us to characterize the proportion of clusters with morphological and dynamical disturbances at high redshift. It is therefore important to examine the impact of the ICM dynamical state on the characterization of the mean pressure profile, which is one of the key elements of cosmological analyses based on tSZ surveys.\\
\indent Hydrodynamical simulations enable us to validate the methods that are used to estimate the ICM properties of galaxy clusters by comparing the results obtained by deprojecting mock observations of synthetic clusters to the actual ICM properties that can be directly extracted from the three-dimensional cube of the simulation. This can help improve the analysis tools developed to estimate the ICM pressure distribution from tSZ observations by characterizing the systematic uncertainties resulting from oversimplified hypotheses and instrumental effects in the deprojection procedure. This paper presents the analysis performed on a sample of clusters from the hydrodynamical simulation \emph{Marenostrum MUltidark SImulations of galaxy Clusters} \citep[MUSIC;][]{sem12}. In this study, we assume that galaxy clusters are spherical objects in hydrostatic equilibrium and we analyze the impact of deviations from spherical symmetry and disturbed ICM regions that can be identified by NIKA2 on the mean pressure profile of a synthetic twin of the NIKA2 cluster sample. This synthetic sample is extracted from the MUSIC simulation in the same region of the mass-redshift plane as that considered for the NIKA2 SZ large program. The simulated cluster sample that we defined helps characterize the analysis methodology used to extract the ICM properties of each cluster of the NIKA2 SZ large program. It will also enable us to study the impact of the different hypotheses underlying the standard definition of the mean pressure profile and the tSZ-mass scaling relation of the sample ({e.g.,} spherical symmetry, hydrostatic equilibrium, self-similarity).\\
\indent This paper is organized as follows. The main characteristics of the NIKA2 instrument and the cluster sample of the NIKA2 SZ large program are described in Sect. \ref{sec:nika2_szlp}. We present the characteristics of the MUSIC simulation and the simulated clusters that are considered for this study in Sect. \ref{sec:music_sample}. We then describe the procedure applied to simulate NIKA2 and \planck\ tSZ maps from MUSIC clusters in Sect. \ref{sec:nika2_simu}. The selection method used to define a synthetic twin of the cluster sample of the NIKA2 SZ large program is also presented. The method used to analyze the simulated tSZ maps to estimate the mean pressure profile associated with the cluster sample is finally presented in Sect. \ref{sec:mean_prof} and the results are interpreted given the known dynamical state of the MUSIC clusters. Finally, we present our conclusions in Sect. \ref{sec:Conclusions}.

\section{NIKA2 SZ large program}\label{sec:nika2_szlp}

\subsection{Thermal Sunyaev-Zel'dovich effect}\label{subsec:tSZ_effect}

In this study, we assume that the total SZ signal measured for each cluster in the NIKA2 SZ large program will be dominated by the tSZ effect and thus use the term "tSZ observations" throughout the paper.

The tSZ effect \citep{sun72,sun80} is due to the inverse Compton scattering of CMB photons by high-energy ICM electrons. It induces a distortion of the CMB blackbody spectrum toward higher frequency and results in an intensity shift relative to the CMB given by
\begin{equation}
        \frac{\Delta I_{tSZ}}{I_0} = y \, f(\nu, T_e),
\label{eq:deltaI}
\end{equation}
where $f(\nu, T_e)$ is the frequency dependence of the tSZ spectrum \citep{bir99,car02}, $T_e$ is the electronic temperature of the ICM, and $y$ is the Compton parameter that characterizes the amplitude of the spectral distortion. This parameter is a dimensionless measure of the line-of-sight integral of the thermal pressure $P_e$ for a given position in the sky, i.e.,
\begin{equation}
        y = \frac{\sigma_{\mathrm{T}}}{m_{e} c^2} \int P_{e} \, dl,
        \label{eq:y_compton}
\end{equation}
where $m_{e}$ is the electron rest mass, $\sigma_{\mathrm{T}}$ the Thomson scattering cross section, and $c$ the speed of light. The spherical integral of the ICM pressure distribution enables us to estimate the integrated Compton parameter $\rm{Y_{tot}}$ that is expected to provide a low-scatter mass proxy for galaxy clusters. In this paper, we consider the spherically integrated Compton parameter up to a cluster radius $\rm{R_{500}}$ for which the mean cluster density is $500$ times the critical density of the Universe.\\
The temperature dependence in the shape of the tSZ spectrum is due to relativistic corrections that induce a shift of the null frequency of the tSZ effect toward higher frequencies \citep{ito98,poi98}. Furthermore, its global effect for frequencies lower than ${\sim}500$~GHz is to decrease the amplitude of the intensity variation induced by the tSZ effect as the ICM electron temperature increases. In this paper, we consider the relativistic corrections to be negligible in comparison to the RMS noise induced by the atmospheric and electronic noise contaminants in the NIKA2 mock observations (see, {e.g.,} \cite{rem18} for details on the importance of relativistic corrections in tSZ analyses). Therefore, we do not take these relativistic corrections into account when we simulate the NIKA2 tSZ maps from the MUSIC data set and for the pressure profile estimation from the simulated maps (see Sect. \ref{sec:nika2_simu} and \ref{sec:mean_prof}).

\subsection{Instrument: the NIKA2 camera}\label{subsec:nika2_cam}

The NIKA2 camera is a continuum instrument installed at the 30 m telescope of IRAM. Three arrays of frequency-multiplexed kinetic inductance detectors \citep[KIDs;][]{mon10,roe12} are installed at the focal plane of the instrument. These arrays enable observing the sky in a field of view of 6.5 arcminutes simultaneously at 150 and 260~GHz. The main beam full width at half maximum (FWHM) measured during the commissioning phase of the camera are 17.7 and 11.2~arcsec at 150 and 260~GHz, respectively. The point source sensitivities of the instrument averaged over various atmospheric conditions and for different sources are $8~\mathrm{mJy.s^{1/2}}$ and $33~\mathrm{mJy.s^{1/2}}$ at 150 and 260~GHz, respectively. Further information on the NIKA2 camera can be found in \cite{ada18}, \cite{cal16}, and \cite{bou16}.\\

The NIKA2 camera is, together with MUSTANG2, AzTEC, and ALMA, one of the only four instruments currently in service that is suitable for high angular resolution tSZ mapping thanks to its resolution, sensitivity, and its capacity to observe in two frequency bands. The MUSTANG2 \citep{dic14} camera installed at the \emph{Green Bank Telescope} has a better angular resolution ($9$~arcsec at 90~GHz) but a smaller field of view ($4$~arcmin). The AzTEC instrument \citep{say18} installed at the \emph{Large Millimeter Telescope} has an angular resolution of $6$~arcsec at 270~GHz and a field of view of $2.4$~arcmin. The Atacama Large Millimeter/submillimeter Array (ALMA) in the compact array configuration reaches an angular resolution of about $2-5$~arcsec for tSZ observations \citep[{e.g.,}][]{kit16}. However, the tSZ signal can only be mapped in limited regions of the observed clusters. The NIKA2 camera is therefore very well suited to map the tSZ signal of high redshift galaxy clusters over a field of view comparable to that of the X-ray observatories considered so far to constrain the pressure profile and scaling relation used in cosmological analyses.

\subsection{Custer sample}\label{subsec:sample_select}

\begin{figure*}[h!]
\centering
\includegraphics[height=6.6cm]{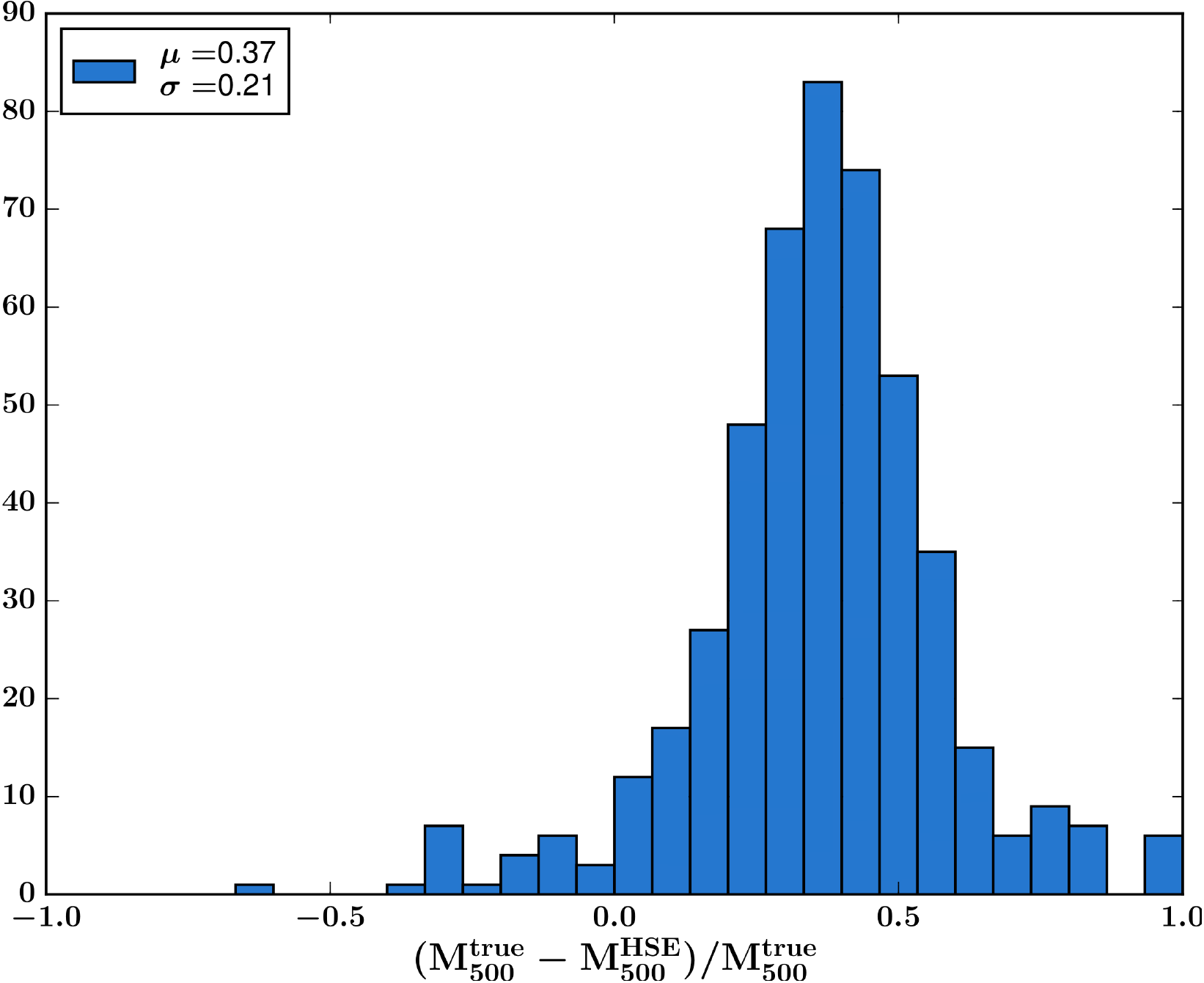}
\hspace{1cm}
\includegraphics[height=6.6cm]{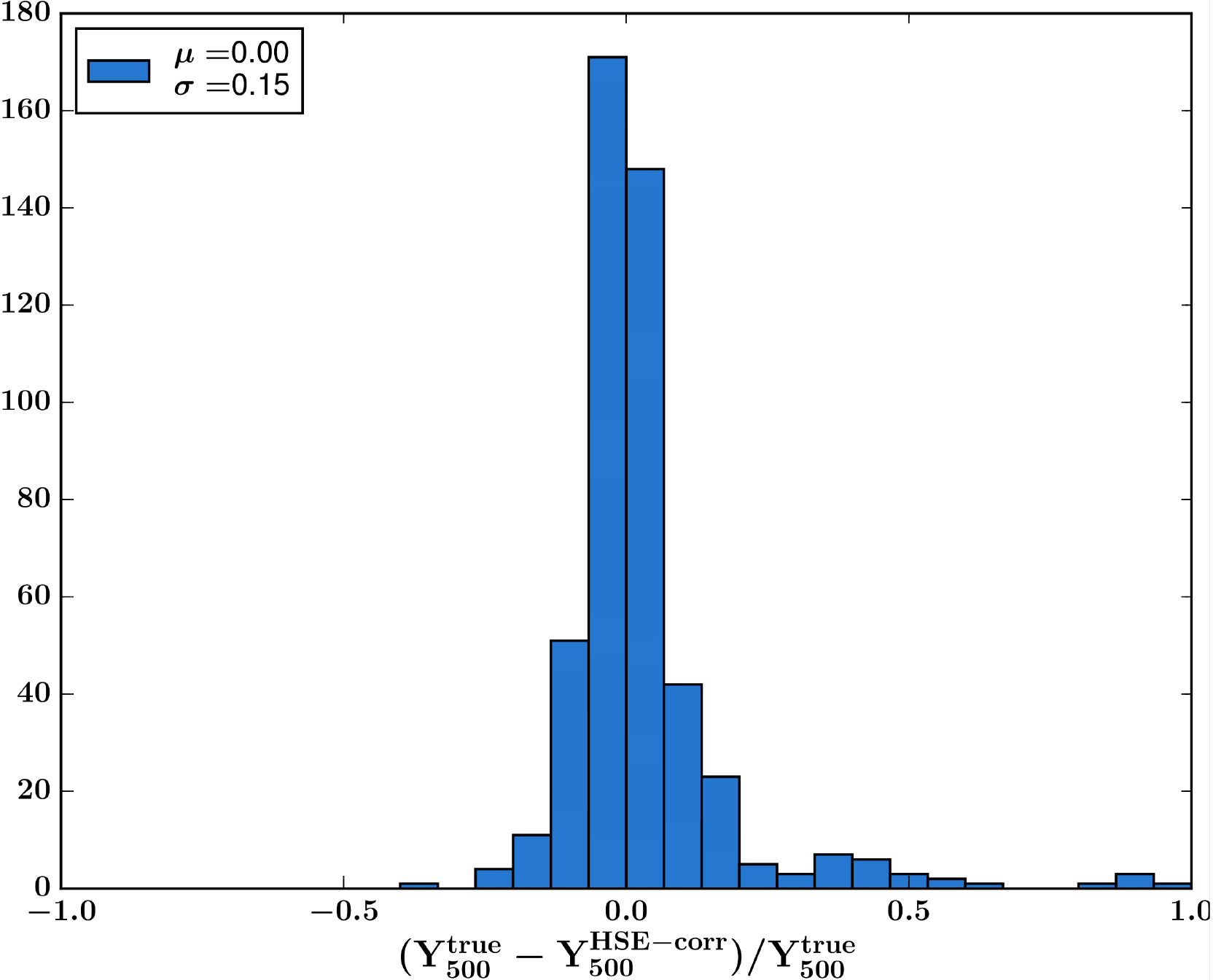}
\caption{{\footnotesize \textbf{Left:} Distribution of the hydrostatic bias values computed for all MUSIC clusters at redshifts $z = 0.54$ and $z = 0.82$. \textbf{Right:} Distribution of the bias on the corrected integrated Compton parameter for the same clusters. The mean and the standard deviation of each histogram are indicated in the upper left corner of the panels.}}
\label{fig:mass_Y500_bias}
\end{figure*}

The selection strategy considered by the NIKA2 collaboration to build the cluster sample for the NIKA2 SZ large program is mainly motivated by the need to select a representative sample of galaxy clusters. The analysis of a representative cluster sample makes it possible to characterize the scaling relation $Y_{500}{-}M_{500}$ and the average pressure profile applicable to the entire population of galaxy clusters regardless of their dynamical state. In addition, the results obtained from the analysis of a representative cluster sample can be used to characterize the global properties of clusters and obtain better control of systematic effects generated by different processes such as merger events with substructures, energy injection into the ICM driven by AGNs or supernovae winds, and gas turbulence. It is however important to note that it is challenging to discriminate among the different origins of ICM disturbances with tSZ observations only. Since the integrated Compton parameter is directly related to the thermal energy content within the clusters (see Sect. \ref{subsec:tSZ_effect}), a selection cut on this parameter meets the requirement of establishing a representative sample, unlike a selection based on the X-ray luminosity, which favors relaxed clusters with a dense core \citep{ros16,ros17,and17}. The representative cluster sample of the NIKA2 SZ large program has been extracted from the tSZ catalogs established by the \planck\ and the Atacama Cosmology Telescope (ACT) collaborations \citep{pla16b,hil18}. Two redshift bins have been considered, i.e., $0.5<z<0.7$ and $0.7<z<0.9$. The selected clusters are distributed uniformly in five mass bins above the selection cut fixed at $M_{500} > 3\times 10^{14}~\mathrm{M_{\odot}}$. More details on the NIKA2 SZ large program cluster sample can be found in \cite{per18}.

\subsection{Mean pressure profile at high redshift}\label{subsec:goal_szlp}

One of the main goals of the ongoing NIKA2 SZ large program is to explore and test the regularity of the pressure profile of galaxy clusters at $z>0.5$. The methodology applied in this program follows the approach used by \cite{arn10} using the \xmm\ observatory to estimate the universal pressure profile from a sample of low redshift clusters at $z<0.2$ and simulations to explore the radial range up to $4R_{500}$. However, using the tSZ effect as a cluster observable allows us to probe directly the pressure distribution in the ICM. In addition, the characterization of the statistical properties of the pressure profiles, in relation to the dynamical state of the selected clusters, brings key information in understanding the selection function of cosmological surveys based on the observation of the tSZ effect and the systematic effects affecting analyses based on the study of the tSZ power spectrum.

Each galaxy cluster in the NIKA2 SZ large program sample has to be analyzed using the same set of hypotheses and procedures to estimate their pressure profile regardless of their morphology or dynamical state. Indeed, as the mean pressure profile of galaxy clusters is used for both the catalog extraction of tSZ surveys \citep[{e.g.,}][]{mel06} and tSZ cosmological analyses based on a mass function \citep[{e.g.,}][]{sal18}, it is essential to preserve the consistency between the halo definition in simulations and that in the NIKA2 SZ large program based on clusters detected in tSZ surveys. It is therefore important to characterize the systematic uncertainties associated with the deprojected pressure profile of galaxy clusters when we consider them as spherical objects in hydrostatic equilibrium, while their morphology and dynamical state can be significantly different from this oversimplified model. The challenge of the NIKA2 SZ large program for cosmology lies in the accurate estimation of the total error budget associated with the pressure profile of each cluster in the sample given their morphology and dynamical state. The procedure used to constrain the mean pressure profile of the galaxy clusters observed by NIKA2 is developed in detail in Sect. \ref{subsec:impact_icm_dist}.

\section{MUSIC-simulated sample of galaxy clusters}\label{sec:music_sample}

Numerical hydrodynamical simulations are valuable tools to study the origin of deviations from the pure gravitational collapse scenario of cluster formation, which predicts a self-similar scaling of the tSZ observable with the cluster mass \citep[see][for a review]{bor11}. Although the integrated tSZ signal has been found to be rather insensitive to the baryonic and sub-grid physics assumed in hydrodynamical simulations \citep[{e.g.,}][]{sha08}, the comparison of observational results related to star formation, radiative cooling, galactic, and AGN feedback with simulations is fundamental to improving our understanding of the poorly known impact of these phenomena on the ICM pressure distribution.\\

The fundamental equations of any numerical hydrodynamical simulation are based on a cosmological model as well as on the laws of gravitation and fluid dynamics \citep[see the review of][for more details on N-body simulation techniques.]{dol08}. The main interest of numerical simulations for cosmology comes from their ability to take into account the nonlinearities inherent to the large structure formation processes in order to estimate the expected shape of the mass function giving the expected abundance of halos as a function of mass and redshift \citep[{e.g.,}][]{tin08}. The mass functions estimated at the end of simulations are therefore assumed to be more realistic than those obtained from the extended Press-Schechter formalism \citep{lac93}. This is why these mass functions are used in current cosmological analyses to constrain cosmological parameters from cluster surveys.\\

Numerical simulations also enable the study of the systematic effects related to known astrophysical processes within galaxy clusters on the characterization of the profiles of their thermodynamic properties and the scaling relations used in cosmological analyses \citep[{e.g.,}][]{pla17}. These studies have become possible because of the increasing mass resolution in the simulations and therefore in their ability to generate processes for which the characteristic scales are smaller than the typical cluster physical scales. This type of analysis is the main goal of the MUSIC simulation \citep{sem12,sem14}. The purpose of this section is to present the fundamental characteristics of the MUSIC simulation in relation with the analysis developed in this paper. This section introduces the basic elements used in Sect. \ref{sec:nika2_simu} and the mean thermodynamic properties of the MUSIC clusters that are considered.

\subsection{Overview of the MUSIC simulation}\label{subsec:music_overview}

The MUSIC data set is based on the cosmological N-body dark matter-only simulation \emph{MultiDark} \citep{pra12}, which has been made by considering a cube of $1 \, h^{-1}~\mathrm{Gpc}$ side length. The MUSIC simulation considers the low-resolution execution of \emph{MultiDark} containing a total of $16.8$ million of dark matter particles. The simulation procedure is based on an adaptive mesh refinement grid initialized to a redshift $z=65$. It considers a standard cosmological model according to the parameters constrained by WMAP7, {i.e.,} $\Omega_m = 0.27$, $\Omega_b = 0.0469$, $\Omega_{\Lambda} = 0.73$, $h=0.7$, $\sigma_8 = 0.82$, and $n = 0.95$ \citep{kom11}.\\

The 283 most massive halos in the \emph{MultiDark} simulation were selected independently of their morphological properties to be simulated again at high resolution by including gas particles. This new simulation is the MUSIC simulation. The selected systems correspond to clusters with a mass enclosed within the viral radius greater than $10^{15}\, h^{-1}~\mathrm{M_{\odot}}$ at $z=0$ and constitute the MUSIC-2 database considered in the analysis developed in this study. This cluster sample represents a large improvement of the statistics compared to the MUSIC-1 database, which contains 164 clusters including 82 systems specifically selected to have the morphological properties of major mergers \citep{for10}. The MUSIC simulation was performed using GADGET ({GAlaxies with Dark matter and Gas intEracT}), a code devoted to hydrodynamic simulations developed by \cite{spr01}. The GADGET code is based on the {smoothed particle hydrodynamics} (SPH) numerical method, particularly suitable for describing physical processes related to fluid dynamics such as the ICM. The MUSIC simulation applies the {zoom-in} technique developed by \cite{kly01}, which uses the particle phase space trajectories of the \emph{MultiDark} simulation to initialize the dynamical quantities of the new particles contained in the 283 selected systems. Each cluster is simulated within a sphere with a radius of $6\, h^{-1}~\mathrm{Mpc}$ at $z=0$ at a resolution such that the masses of dark matter and gas particles are given by $m_{\mathrm{DM}} = 9.0\times 10^8\, h^{-1}~\mathrm{M_{\odot}}$ and $m_{\mathrm{g}} = 1.9\times 10^8\, h^{-1}~\mathrm{M_{\odot}}$, respectively. The MUSIC clusters are therefore resimulated improving the original \emph{MultiDark} resolution by a factor of 8. The regions adjacent to each sphere are also simulated with a decreasing resolution with radius until the resolution of the \emph{MultiDark} simulation is reached. This enabled us to take into account the physical processes located in the periphery of the MUSIC clusters while optimizing the computing time allocated to the simulation. In addition to the standard equations of gravitation and fluid dynamics, the MUSIC simulation includes additional physical processes such as gas cooling by thermal emission, stellar formation, and energy injection into the environment by supernovae winds \citep{spr03}. A total of 15 instant captures of the properties of each cluster are made between the redshifts $z=9$ and $z=0$. These snapshots enabled us to trace the dynamical behavior of the MUSIC clusters over the relevant redshift interval to study the galaxy cluster formation processes.\\

The analysis developed in this paper relies on the MUSIC snapshots made at redshifts $z=0.54$ and $z=0.82$ to study the thermodynamic properties of the MUSIC clusters in a range of redshifts that is similar to those considered for NIKA2 SZ large program. The MUSIC clusters at redshifts $z=0.54$ and $z=0.82$, which are selected in Sect. \ref{sec:nika2_simu} to build a twin sample of the NIKA2 SZ large program sample are thus called members of bins 1 and 2, respectively.

\subsection{MUSIC \emph{y}-maps and three-dimensional pressure profiles}\label{subsec:music_products}

\begin{figure*}[h!]
\centering
\includegraphics[height=6.6cm]{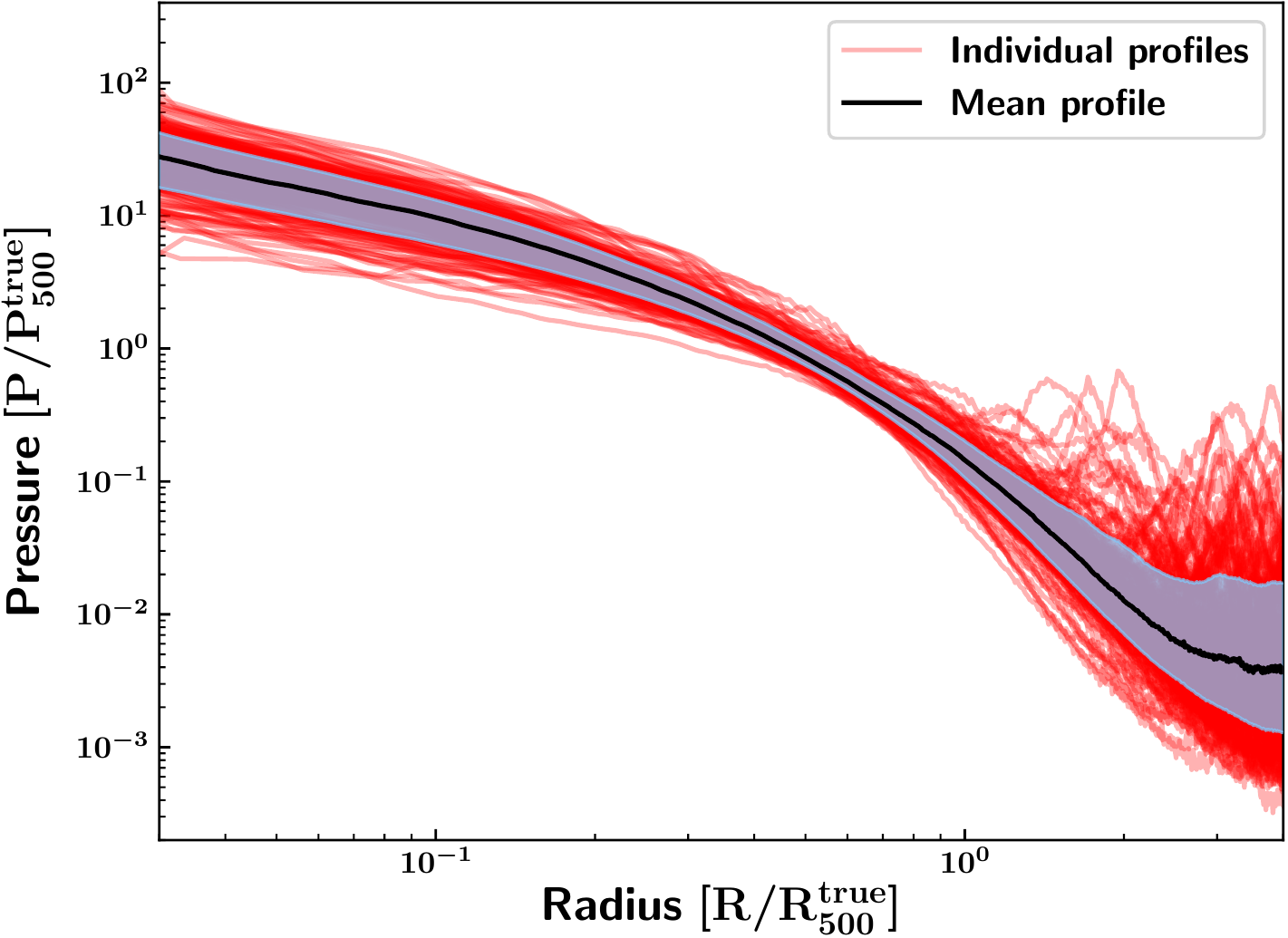}
\hspace{0.4cm}
\includegraphics[height=6.6cm]{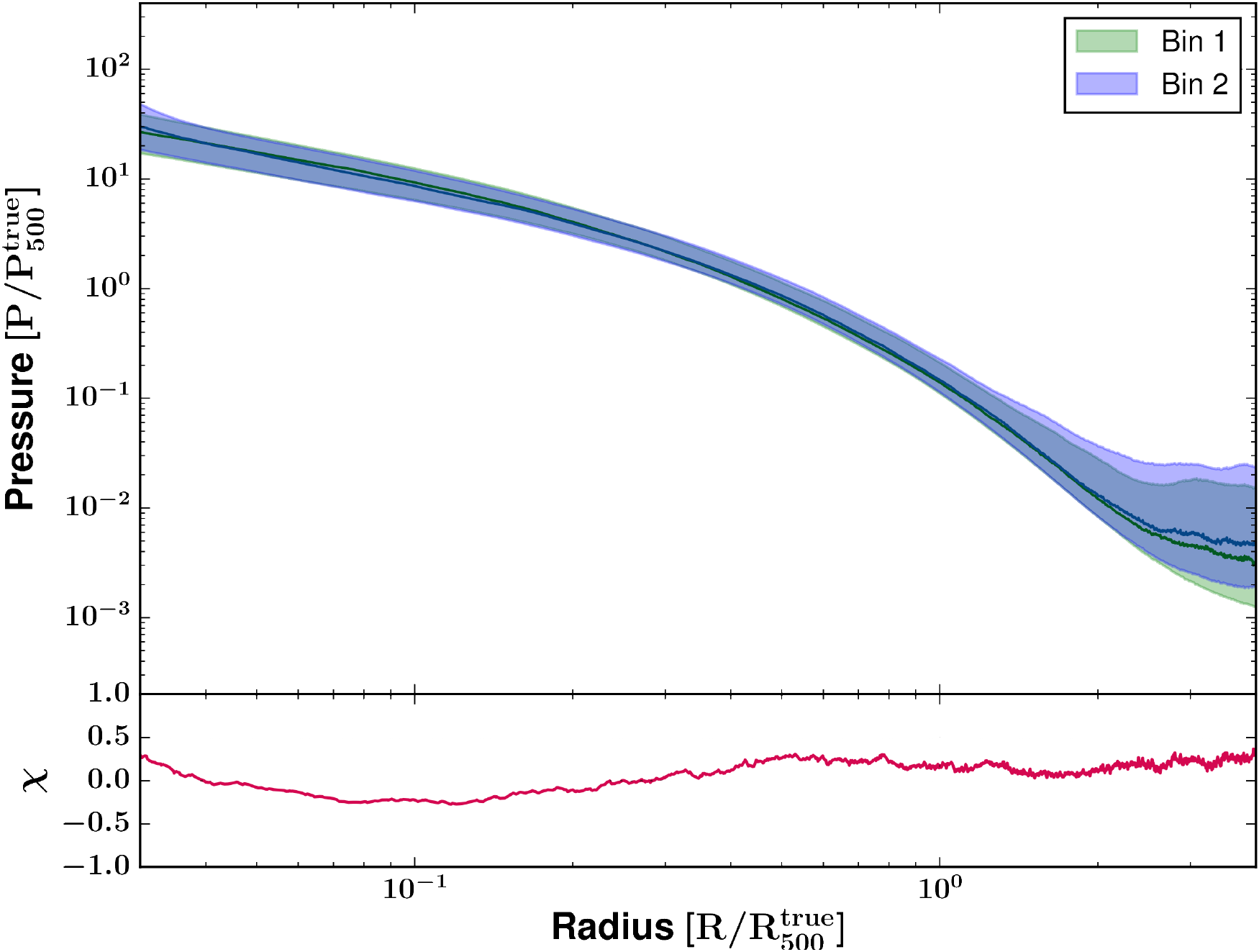}
\caption{{\footnotesize \textbf{Left:} Normalized pressure profiles extracted from the MUSIC simulation for all the clusters located at a redshift $z=0.54$ (red lines). The mean pressure profile is represented in black and the intrinsic scatter of the distribution at $1\sigma$ is given by the blue region. \textbf{Right:} Comparison of the mean normalized pressure profiles estimated from the MUSIC profiles extracted at $z=0.54$ (green), and at $z=0.82$ (blue). The difference between the two profiles divided by the mean of their associated uncertainties is shown in red in the lower panel.}}
\label{fig:MUSIC_P_profiles}
\end{figure*}

In order to carry out an analysis of the impact of the ICM disturbances that can be identified by NIKA2 on the estimation of the pressure profile at high redshift, it is first necessary to obtain Compton parameter maps and thermodynamic profiles associated with each simulated cluster.\\

Compton parameter maps are computed by considering a single line of sight for each cluster. The pressure associated with the gas particles contained in a cylinder is integrated along its axis, aligned with the considered line of sight. The length of the cylinder corresponds to six times the virial radius $R_{\mathrm{vir}}$ of the considered cluster and its base has a radius equal to $3R_{\mathrm{vir}} \simeq 5R_{500}$. We defined all the MUSIC Compton parameter maps using a square grid of 10~Mpc on each side with a pixel size of 10~kpc. At the redshifts considered for this study, this corresponds to a field of view of about 23~arcmin wide and a pixel resolution of about 1.4~arcsec. These maps are therefore well adapted to the instrumental characteristics of NIKA2 because the spatial distribution of the tSZ signal is defined over the entire NIKA2 field of view and the pixel size is small enough compared to the beam extension of NIKA2 at 150~GHz. We used the MUSIC Compton parameter maps in the method developed in Sect. \ref{sec:nika2_simu} to simulate the NIKA2 and \planck\ tSZ maps of the selected MUSIC clusters.\\

It is also necessary to obtain the thermodynamic profiles of the MUSIC clusters to constrain the impact of ICM disturbances on the pressure profile estimated by deprojection of the tSZ signal from the simulated maps. In this paper, the cluster centers considered for the profile extraction from the simulation cube and from the simulated $y$-maps (see Sect. \ref{subsec:profile_extract}) were fixed to the positions obtained from the halo definitions in the \emph{MultiDark} simulation to maintain the consistency between the definitions of halo centers. The radial distributions of the ICM pressure, density, temperature, entropy, and the cluster total mass were calculated for each cluster at the two redshifts considered by averaging the values of the thermodynamic quantities associated with the particles of the simulation in concentric spherical shells. The total sphere considered has a radius equal to $3R_{\mathrm{vir}}$ and the binning of the spherical shells is 10~kpc. The uncertainty associated with each pressure point corresponds to the error on the mean of the pressure values contained in the considered spherical shells. The uncertainties on the profiles increase in the cluster outskirts because of the increase in the standard deviation of the thermodynamic quantities caused by deviations from spherical symmetry and hydrostatic equilibrium in these regions. Moreover, it should be noted that the decrease in the resolution of the simulation in the periphery of the clusters (see Sect. \ref{subsec:music_overview}) implies that the number of particles considered to average the thermodynamic quantities is also reduced. The pressure profiles extracted directly from the simulation are nevertheless very well constrained from the center of the clusters to radii on the order of $3R_{500}$. We thus compared these profiles in Sect. \ref{subsec:impact_icm_dist} to the pressure distribution estimated by deprojection of the tSZ signal in the simulated NIKA2 maps.

\subsection{Integrated parameters of the MUSIC synthetic clusters}\label{subsec:music_integ}

To determine the intrinsic dynamical properties of the MUSIC clusters, we estimated the mass and integrated Compton parameter of each cluster at the two considered redshifts. The quantities $M_{500}$ and $Y_{500}$ represent the fundamental quantities for cosmological analyses. The mass $M_{500}$ is defined from the radius $R_{500}$ using the following relation:
\begin{equation}
M_{500} = \frac{4}{3}\pi R_{500}^3 \times 500\rho_c
\label{eq:M500MUSIC}
.\end{equation}
The integrated Compton parameter $Y_{500}$ is obtained by the spherical integral of the pressure profile up to $R_{500}$. The $R_{500}$ radius is therefore an essential quantity to define the integrated parameters of the clusters. The mass profile $\mathrm{M}(r)$ of each simulated cluster is used to calculate a mean matter density profile $\langle \rho \rangle (r)$ using the volume profile $V(r) = \frac{4}{3}\pi r^3$ to define the radius $R_{500}$ and therefore the quantities $M_{500}$ and $Y_{500}$. We used two different procedures to estimate the mass profile $\mathrm{M}(r)$ of the simulated clusters.\\
\indent The first procedure is based on the total mass profile $\mathrm{M^{true}}(r)$ extracted from the MUSIC simulation by computing the total mass induced by dark matter and gas particles inside spheres of increasing radius. The radius $R_{500}^{\mathrm{true}}$ is estimated from the mean matter density profile obtained with this mass profile. The equation (\ref{eq:M500MUSIC}) is used to calculate $M_{500}^{\mathrm{true}}$. The pressure profile extracted from the simulation (see Sect. \ref{subsec:music_products}) is integrated up to $R_{500}^{\mathrm{true}}$ to estimate $Y_{500}^{\mathrm{true}}$.\\

The second procedure is motivated by the fact that the observations made by NIKA2 for its SZ large program cannot be used to estimate the radius $R_{500}^{\mathrm{true}}$. Indeed, we will not have access to unbiased mass measurements for the clusters of the large program. A possible way to compute the mass profile with NIKA2 consist in using the assumption of hydrostatic equilibrium and considering the estimated pressure profile and a gas density profile deprojected from X-ray observations. The advantage of this approach is that we do not need to consider a previously calibrated scaling relation or assumptions on the gas mass fraction to estimate the mass of the considered clusters. We thus calculated for each MUSIC synthetic cluster the hydrostatic mass profile $\mathrm{M^{HSE}}(r)$ assuming spherical symmetry by combining their pressure $P_e(r)$ and density $n_e(r)$ profiles using the following equation:
\begin{equation}
\mathrm{M^{HSE}}(r) = -\frac{r^2}{G\mu m_p n_e(r)} \times \frac{d \, P_e(r)}{dr}
\label{eq:mass_HSE}
,\end{equation}
where $m_p$ is the proton mass, $\mu$ is the mean molecular weight fixed to a value of 0.62\footnote{Mass of the ICM particles in proton mass units ($m = \mu m_p$)}, and $G$ is the gravitational constant. As of the $\mathrm{M^{true}}(r)$ mass profile, this mass profile is used to estimate the value of $R_{500}^{\mathrm{HSE}}$. The latter allowed us to define the integrated quantities $M_{500}^{\mathrm{HSE}}$ and $Y_{500}^{\mathrm{HSE}}$ associated with each MUSIC cluster.\\

The estimates of the true total mass of the simulated clusters and their mass calculated under the hydrostatic equilibrium hypothesis make it possible to constrain the value of the hydrostatic bias for each MUSIC cluster at an overdensity equal to 500, i.e.,
\begin{equation}
b = \frac{M_{500}^{\mathrm{true}} - M_{500}^{\mathrm{HSE}}}{M_{500}^{\mathrm{true}}}
.\end{equation}
The mean of the $b$ values obtained at the two considered redshifts are compatible\footnote{We observe a relative difference of 4\% between the mean hydrostatic biases estimated at the two redshifts.}. The left panel of Fig. \ref{fig:mass_Y500_bias} represents the distribution of the $b$ values for all the MUSIC clusters at $z=0.54$ and $z=0.82$. The mean hydrostatic bias associated with the MUSIC clusters is given by $\mu_b = 0.37$, which is compatible with the observations made by the project \emph{Weighing the Giants} \citep{lin14}, although higher than most current constraints between 0.1 and 0.3 \citep[see, {e.g.,} Fig. 10 in][]{sal18}. We also observe on the histogram of the $b$ values obtained in the simulation that the hydrostatic bias is subject to a large scatter: $\sigma_b = 0.21$. Furthermore, the extreme values of the bias are mainly associated with major mergers for which the hydrostatic equilibrium hypothesis is not reliable. This is discussed later in Sect. \ref{subsec:nika2_sample}.\\ 
\begin{figure*}[h!]
\centering
\includegraphics[height=6.6cm]{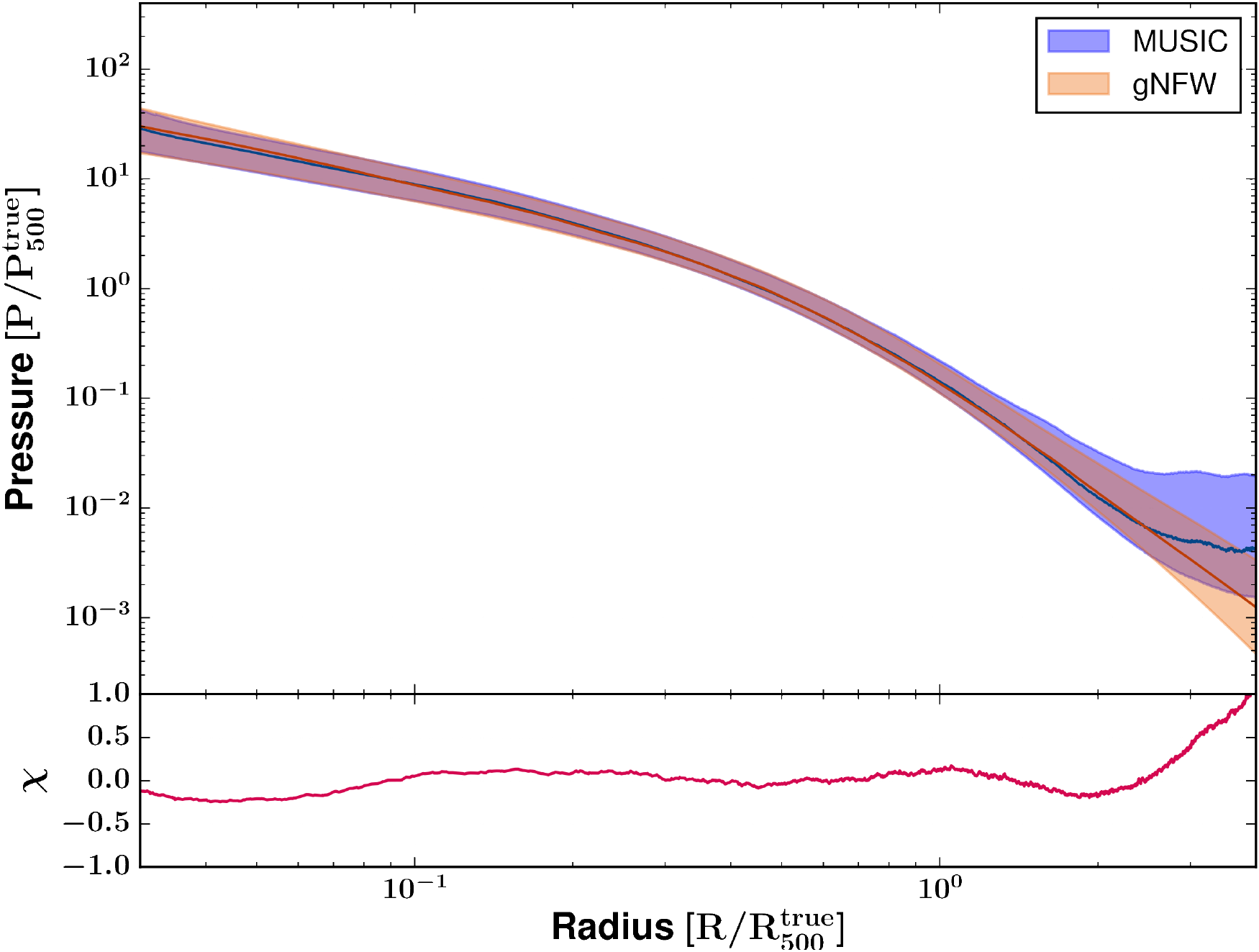}
\hspace{0.6cm}
\includegraphics[height=6.6cm]{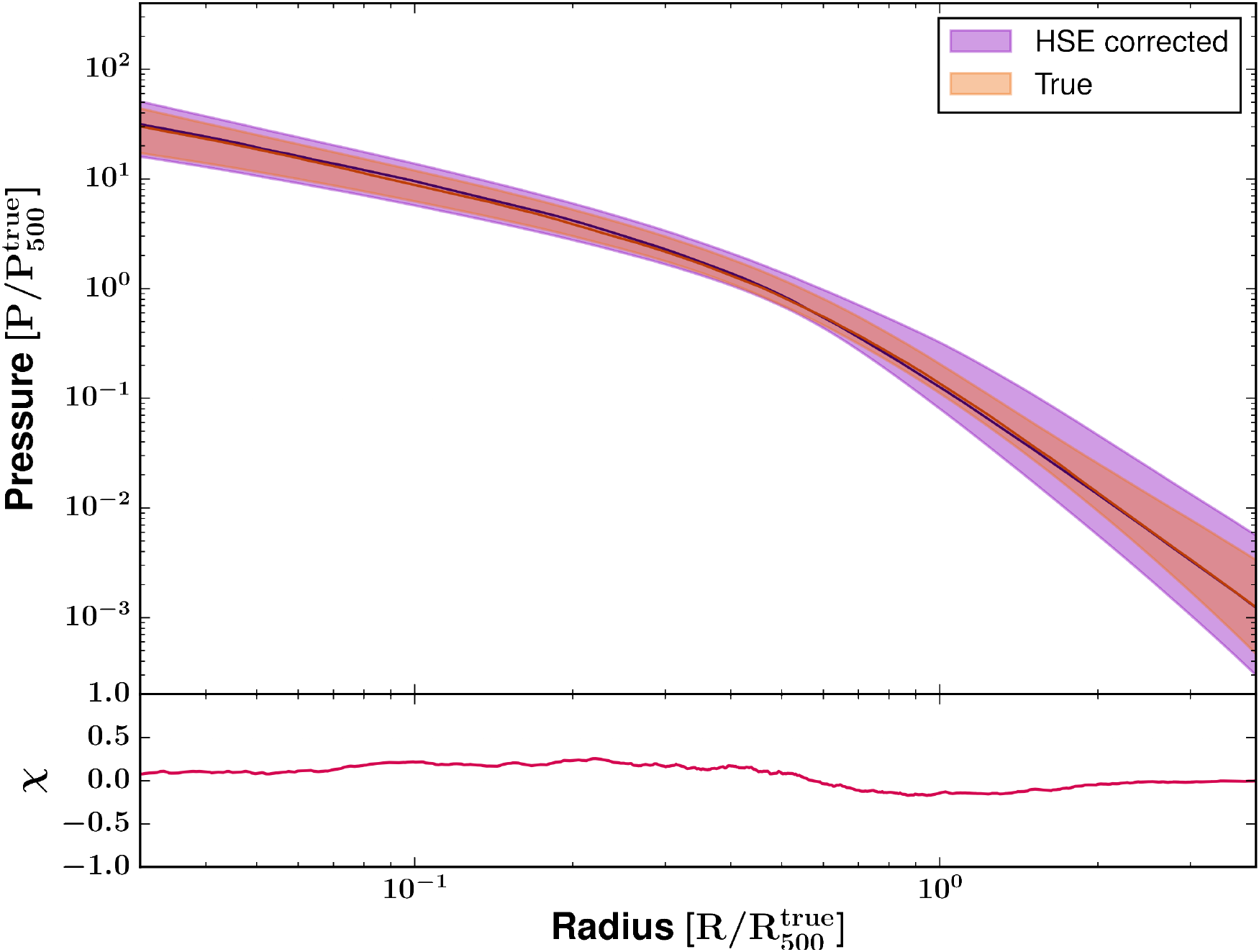}
\caption{{\footnotesize \textbf{Left:} Mean normalized pressure profiles estimated by considering the profiles extracted from the MUSIC simulation (blue) and the gNFW fits of these profiles (orange) for each cluster at $z=0.54$ and $z=0.82$. \textbf{Right:} Mean pressure profiles obtained by normalizing the gNFW profiles associated with each MUSIC cluster by the values of $R_{500}^{\mathrm{true}}$ and $P_{500}^{\mathrm{true}}$ (orange) and by the values of $R_{500}^{\mathrm{HSE-corr}}$ and $P_{500}^{\mathrm{HSE-corr}}$ obtained after correction of the hydrostatic mass of each cluster by the mean hydrostatic bias of the MUSIC simulation at the two considered redshifts (magenta). The differences between the two profiles divided by the mean of their associated uncertainties are shown in red in each lower panel.}}
\label{fig:MUSIC_gNFW_HSE}
\end{figure*}

The hydrostatic bias associated with each cluster of the NIKA2 SZ large program is not known a priori. The hydrostatic mass $M_{500}^{\mathrm{HSE}}$ is thus usually corrected using the mean hydrostatic bias from the results of other studies. In the case studied in this paper, we estimated the corrected hydrostatic mass of each MUSIC cluster using the following relation:
\begin{equation}
M_{500}^{\mathrm{HSE-corr}} = M_{500}^{\mathrm{HSE}} / (1 - \mu_b)
.\end{equation}
The correction of the hydrostatic mass by the mean hydrostatic bias makes it possible to cancel on average the bias observed between the mass estimates but has no effect on the dispersion $\sigma_b$ of the distribution. Using the mass $M_{500}^{\mathrm{HSE-corr}}$ allowed us to define a value of $R_{500}^{\mathrm{HSE-corr}}$ by applying the equation (\ref{eq:M500MUSIC}) that can be used as an outer boundary in the integration of the pressure profile to estimate the integrated Compton parameter $Y_{500}^{\mathrm{HSE-corr}}$ for each cluster. The right panel of Fig. \ref{fig:mass_Y500_bias} shows the histogram of the bias on the $Y_{500}$ measurement made by considering the estimate of the $R_{500}^{\mathrm{HSE-corr}}$ radius as an outer boundary of the integral of the pressure profile of each cluster instead of the $R_{500}^{\mathrm{true}}$ radius. The histogram obtained is centered on 0 but has a standard deviation of 15\%. The use of the hydrostatic equilibrium hypothesis and a mean hydrostatic bias thus induces an additional scatter over all the measured integrated quantities. The $Y_{500}{-}M_{500}$ scaling relation calibrated using X-ray and tSZ measurements therefore has an increased dispersion compared to its intrinsic dispersion owing to the use of the hydrostatic equilibrium hypothesis. It is important to take this systematic effect into account when the tSZ-mass scaling relation associated with the sample of the NIKA2 SZ large program is estimated.

\subsection{Mean pressure profile of the MUSIC synthetic sample}\label{subsec:music_prof}

The objective of this section is to estimate the mean pressure profile of the MUSIC clusters based on the profiles extracted directly from the simulation. The uncertainty associated with the mean pressure profile also takes into account the number of profiles considered for its estimation and can be propagated in a cosmological analysis based on a tSZ survey up to the constraints on the cosmological parameters. However, we are more interested in the intrinsic scatter associated with the mean pressure profile of a cluster sample in this study because it corresponds to the potential systematic error made if the self-similarity hypothesis cannot be applied to the whole cluster population.\\

We estimated the mean pressure profiles of the MUSIC clusters by considering all their associated profiles at $z=0.54$ and $z=0.82$. The left panel of Fig. \ref{fig:MUSIC_P_profiles} represents all pressure profiles (red curves) of the MUSIC synthetic clusters at redshift $z=0.54$ after normalization of the radius by the value of $R_{500}^{\mathrm{true}}$ associated with each cluster and the pressure by the amplitude factor $P_{500}^{\mathrm{true}}$. This factor gives the scaling relation between the pressure content and the cluster total mass in the self-similar model \citep{arn10}
\begin{equation}
P_{500}^{\mathrm{true}} = 1.65 \times 10^{-3} \, E_z^{8/3} \, \left[ \frac{M_{500}^{\mathrm{true}}}{3 \times 10^{14} \, h_{70}^{-1}~\mathrm{M_{\odot}}}\right]^{2/3} \, h_{70}^2~\mathrm{keV} \, \mathrm{cm^{-3}}
,\end{equation}
where $E_z$ is the ratio of the Hubble constant at redshift $z$ to its present value $H_0$, and $h_{70} = H_0 / [70~\mathrm{km/s/Mpc}]$. The normalized pressure distribution is modeled at each radius by an asymmetric lognormal distribution\footnote{The logarithm of the pressure is modeled by a Gaussian with two distinct standard deviations on either side of the distribution peak.} to constrain the mean pressure profile and its intrinsic dispersion. The computed profile is represented by a black line on the left panel of Fig. \ref{fig:MUSIC_P_profiles} and the standard deviation at $1\sigma$ of the distribution of the normalized pressure profiles is given by the blue region. The mean pressure profile shows a flattening for radii $r>2R_{500}$ around the splashback radius \citep{mor15}. In addition, the dispersion of the profile distribution increases significantly in the same radius range. This is because of the deviations from the gas dynamic equilibrium at radii larger than the virial radius. Accretion of the surrounding environment and the presence of dense substructures and gas shocks in these regions lead to an increase in the mean thermal pressure and fluctuations of the latter, which are responsible for the flattening of the mean pressure profile and the increase of its associated scatter, respectively.\\
\begin{figure*}[h!]
\centering
\includegraphics[height=6.4cm]{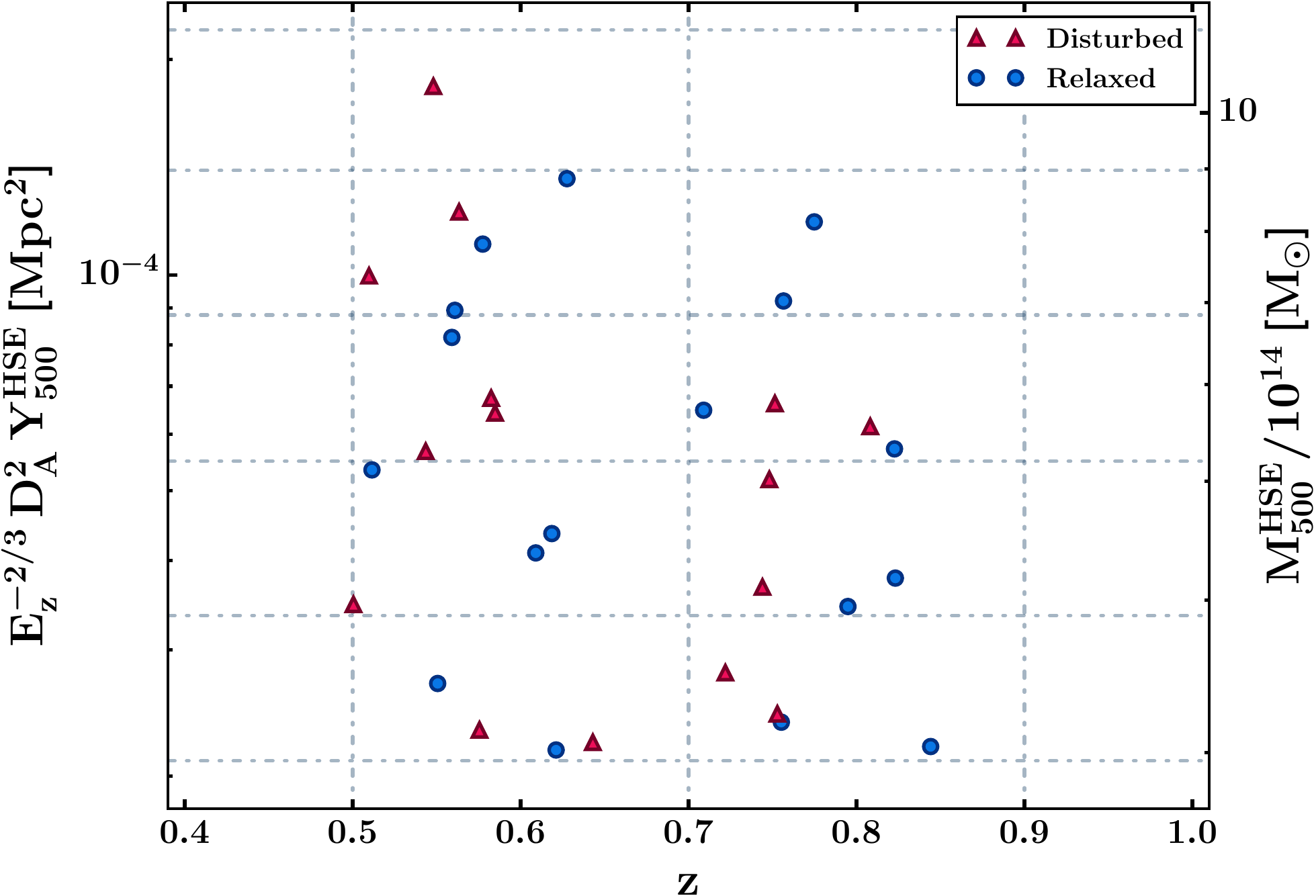}
\hspace{0.6cm}
\includegraphics[height=6.4cm]{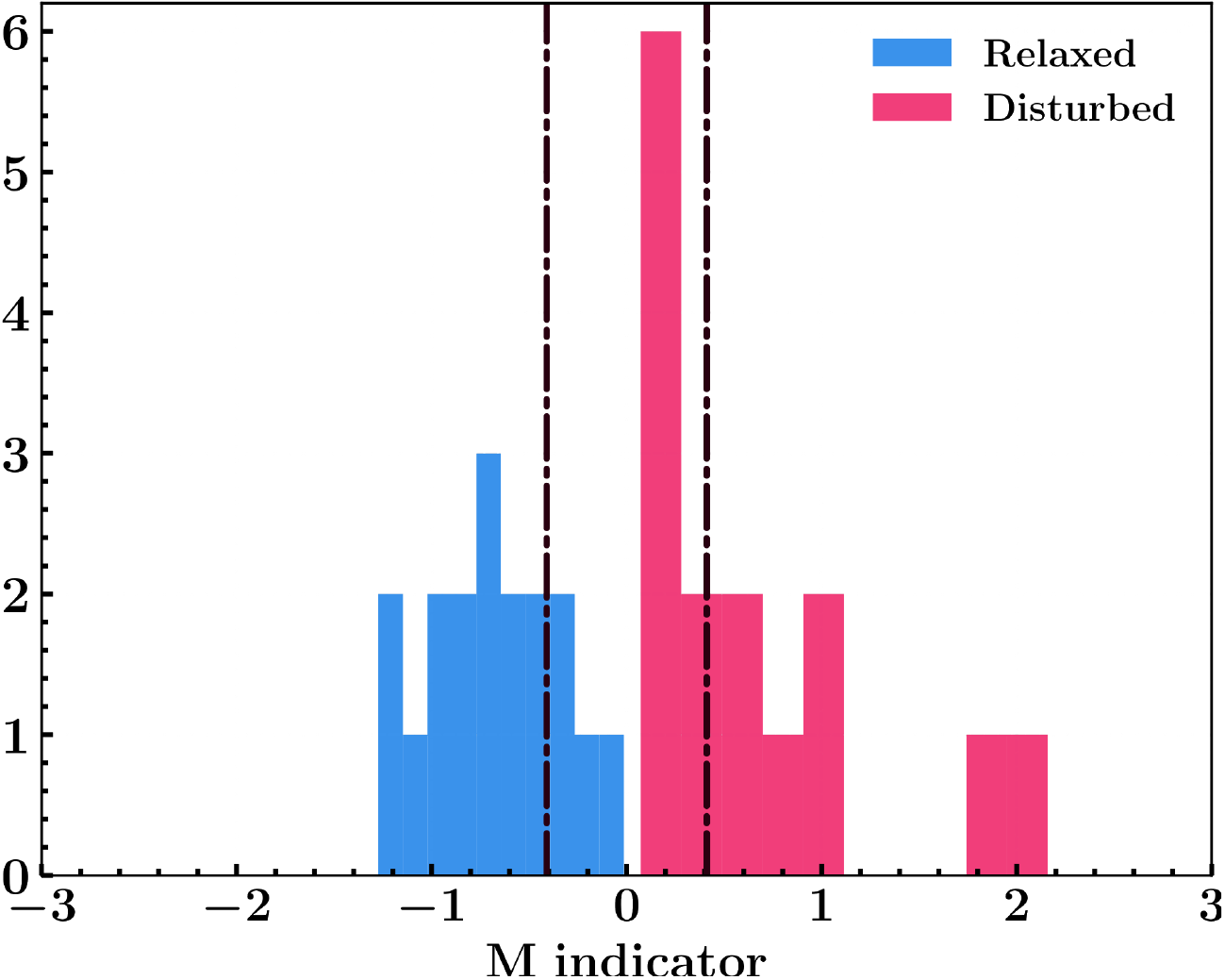}
\caption{{\footnotesize \textbf{Left:} Distribution of the selected MUSIC clusters in the mass-redshift plane. The mass and redshift bins considered are identical to those of NIKA2 SZ large program. The redshifts in each bins are generated randomly using a uniform distribution within their respective boundaries for display reasons. Morphologically relaxed and disturbed clusters are indicated by blue dots and red triangles, respectively. \textbf{Right:} Distributions of the morphological indicators computed for the selected MUSIC clusters for the relaxed (blue) and disturbed (red) subsamples. The vertical dash-dotted lines enclose the population of clusters with a hybrid dynamical state (see Sect. \ref{subsec:nika2_sample}).}}
\label{fig:MUSIC_NIKA2_sample}
\end{figure*}

We performed an identical analysis for the pressure profiles associated with the MUSIC clusters at redshift $z=0.82$. The mean pressure profiles obtained at the two considered redshifts are represented in blue and green on the right panel of Fig. \ref{fig:MUSIC_P_profiles}. The red curve shown in the lower panel of the figure corresponds to the difference between the two profiles divided by the mean dispersion observed at each radius. We note that no significant variation in the shape and amplitude of the mean pressure profile is observed between the redshifts $z=0.54$ and $z=0.82$ of the MUSIC simulation. We therefore chose to consider a mean pressure profile combining all the profiles estimated at these two redshifts in the analysis developed in Sect. \ref{subsec:impact_icm_dist}.\\

In order to reduce the analysis time associated with the estimation of the pressure profile of each cluster considered in the study developed in Sect. \ref{subsec:profile_extract}, we chose to model the pressure distribution in the ICM by a parametric generalized Navarro-Frenk-White profile \citep[gNFW; ][]{nag07},
\begin{equation}
        P_e(r) = \frac{P_0}{\left(\frac{r}{r_p}\right)^c \left(1+\left(\frac{r}{r_p}\right)^a\right)^{\frac{b-c}{a}}},
\label{eq:gNFW}
\end{equation}
where $P_0$ is a normalization constant, $r_p$ is a characteristic radius, and $a$ characterizes the size of the transition between the two profile slopes $b$ and $c$ at large and small radii, respectively. This model is therefore defined by only five free parameters while a nonparametric profile would contain more than twice as many constrained values for typical observations with NIKA2 \citep{rup18}. It is therefore necessary to check whether the gNFW model is appropriate to describe the mean pressure profile of the MUSIC clusters at the considered redshifts. Each pressure profile extracted from the MUSIC simulation at both redshifts is therefore fitted by a gNFW model and normalized by the corresponding values of $R_{500}^{\mathrm{true}}$ and $P_{500}^{\mathrm{true}}$. The resulting distribution of gNFW profiles is used to estimate the mean pressure profile and its associated scatter. In Fig. \ref{fig:MUSIC_gNFW_HSE}, the mean pressure profile estimated using the gNFW model (orange line) is compared to the mean profile obtained by considering the pressure profiles extracted directly from the simulation (blue line). The weighted difference between the profiles is shown in the lower panel (red line). We do not observe any significant discrepancy between the two profiles for radii $r<3R_{500}$. In addition, the difference caused by the flattening of the pressure profiles extracted from the simulation observed between the mean profiles for radii greater than $3R_{500}$ results in a systematic bias on the integrated Compton parameter $Y_{\mathrm{5R500}}$ of about 6\%. This difference is four times smaller than the typical relative uncertainty on the integrated tSZ signal measured by \planck\ \citep{pla16b}. The gNFW profile is therefore well adapted to model the pressure distribution of the MUSIC clusters from a combined NIKA2/\planck\ analysis (see Sect. \ref{subsec:profile_extract}).\\
\begin{figure*}[h!]
\centering
\includegraphics[height=6.6cm]{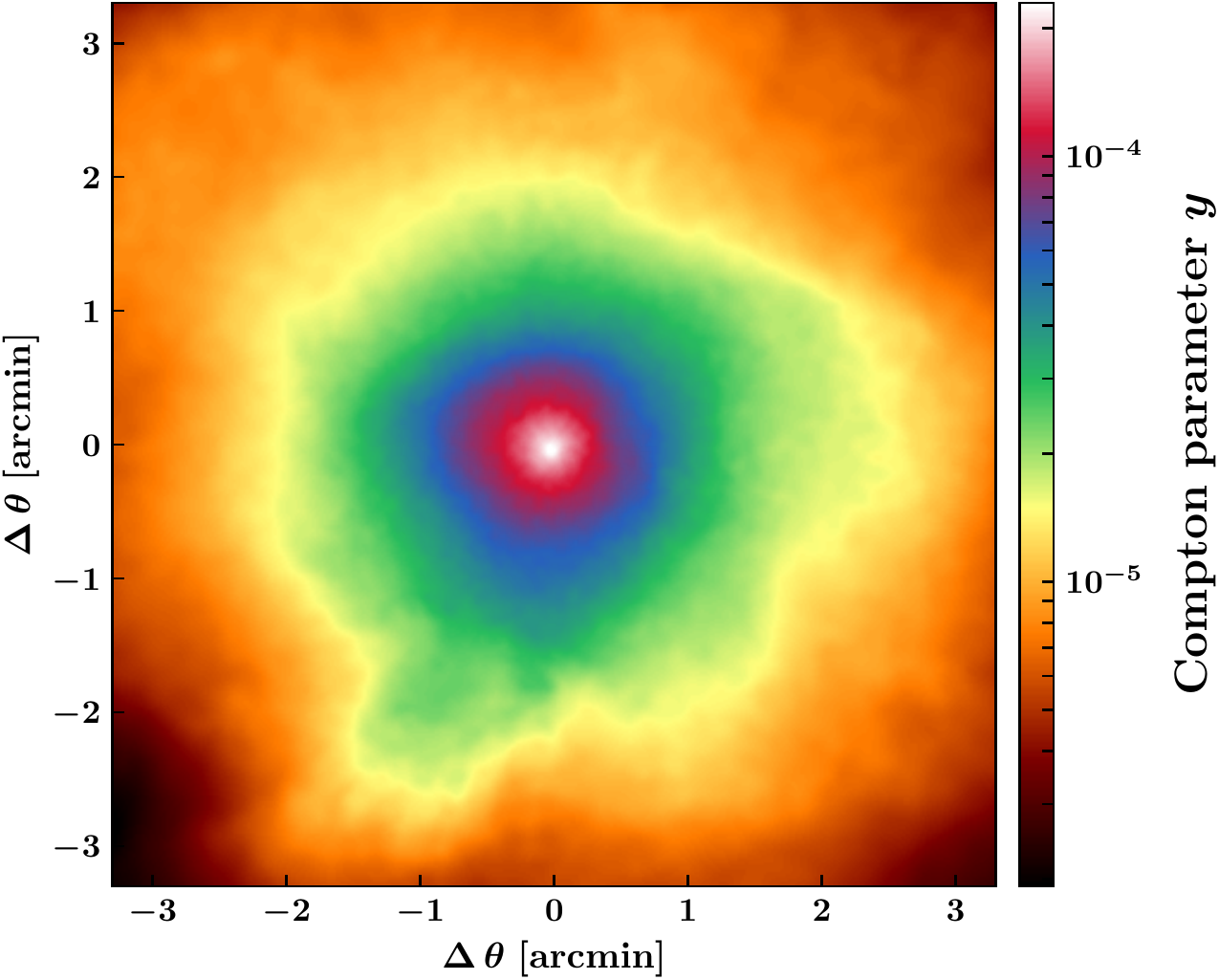}
\hspace{0.6cm}
\includegraphics[height=6.6cm]{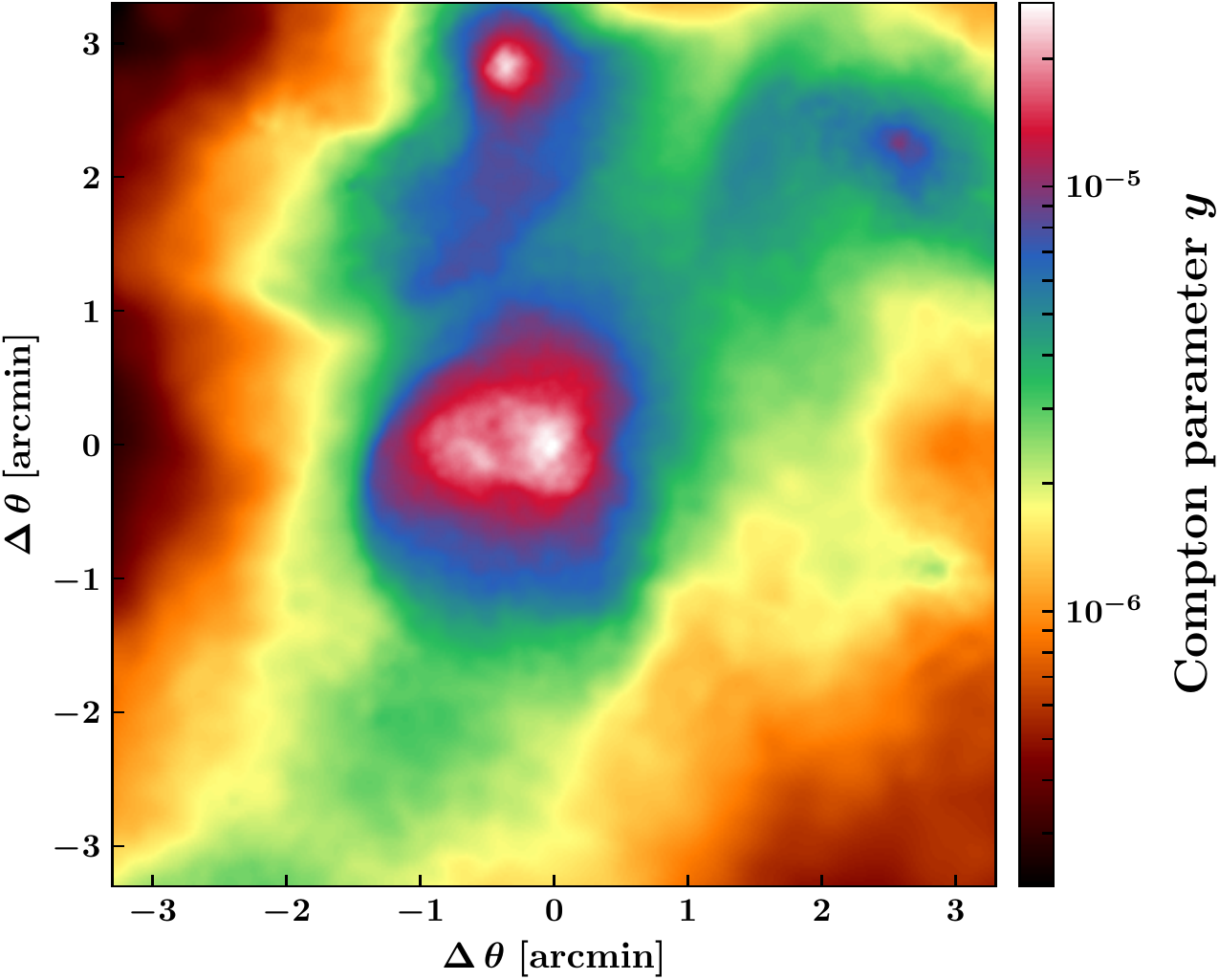}
\caption{{\footnotesize \textbf{Left:} Compton parameter map of the MUSIC cluster $\mathrm{n}^{\circ}7$ at redshift $z=0.54$ considered as a relaxed system in the definition of the simulated cluster sample in Sect. \ref{subsec:simu_obs}.  \textbf{Right:} Compton parameter map of the MUSIC cluster $\mathrm{n}^{\circ}130$ at redshift $z=0.54$ considered as a disturbed system. This represents an extreme case in which both individual substructures and an extension of the ICM of the main halo are clearly identified.}}
\label{fig:def_relax_disturbed}
\end{figure*}

Using the hydrostatic equilibrium hypothesis also has an effect on the mean pressure profile estimated from tSZ observations. As shown in Sect. \ref{subsec:music_integ} the value of $R_{500}^{\mathrm{HSE-corr}}$ estimated from the total mass, under the assumption of hydrostatic equilibrium, corrected by the mean hydrostatic bias, is not biased but is scattered around the value of $R_{500}^{\mathrm{true}}$. The same conclusion is made if we consider the normalization coefficient $P_{500}^{\mathrm{HSE-corr}}$. As shown in the right panel of Fig. \ref{fig:MUSIC_gNFW_HSE}, the mean pressure profile obtained under the hydrostatic equilibrium assumption by correcting the integrated quantities by the mean hydrostatic bias (magenta curve) is identical to the mean pressure profile obtained by considering the values of $R_{500}^{\mathrm{true}}$ and $P_{500}^{\mathrm{true}}$ for the normalization (orange curve). However, its associated scatter is 5 to 60\% higher for radii greater than $0.5R_{500}$. The estimation of the intrinsic scatter of the distribution of pressure profiles at high redshift is important for cosmological analyses as it traces the potential systematic uncertainty associated with the cluster self-similarity assumption. It is therefore important to investigate whether this increase in dispersion associated with the mean pressure profile, caused by the use of the hydrostatic equilibrium hypothesis, is comparable to the additional dispersion induced by systematic effects due to disturbances of the ICM on the reconstruction of the pressure profile from NIKA2 tSZ observations. Understanding the role of the different processes responsible for the dispersion of pressure profiles estimated from tSZ observations will allow us to establish methods that take these systematic effects into account in future cosmological analyses.

\section{Simulation of NIKA2 SZ large program observations}\label{sec:nika2_simu}

In this section, we aim to present the method used to simulate NIKA2 and \planck\ tSZ maps of a sample of MUSIC clusters similar to the sample considered for the NIKA2 SZ large program. These maps constitute the data set used in the analysis developed in Sect. \ref{sec:mean_prof} to study the effect of ICM perturbations on the estimation of the mean pressure profile using the NIKA2 tSZ analysis pipeline.

\subsection{Definition of the synthetic cluster sample}\label{subsec:nika2_sample}

The galaxy cluster selection procedure used for the NIKA2 SZ large program (see Sect. \ref{subsec:sample_select}) was applied to establish a synthetic sample based on the MUSIC data. In order to reduce the analysis time required in the study developed in Sect. \ref{subsec:profile_extract} without increasing significantly the statistical uncertainties on the estimation of the mean pressure profile, we chose to populate each mass and redshift bin considered for the NIKA2 SZ large program by four MUSIC clusters instead of five. The number of clusters selected is thus comparable to the number considered for the NIKA2 SZ large program. The clusters of the synthetic sample are presented in the mass-redshift plane in the left panel of Fig. \ref{fig:MUSIC_NIKA2_sample}. All the simulated clusters of bin 1 are  located at a redshift $z=0.54$ and those of bin 2 at a redshift $z=0.82$. However, only for display reasons, each cluster has a randomly generated redshift assuming a uniform distribution within their respective bins. Since the cluster selection procedure of the NIKA2 SZ large program was carried out by applying a mass cut in the \planck\ and ACT catalogs (see Sect. \ref{subsec:sample_select}), we used the hydrostatic mass estimates of the MUSIC clusters $M_{500}^{\mathrm{HSE}}$ to select the MUSIC clusters from the whole simulated sample. We only considered MUSIC clusters with a hydrostatic mass greater than $3\times 10^{14}~\mathrm{M_{\odot}}$ to populate the considered mass and redshift bins. This cut reduces the size of the initial MUSIC sample to a total of 82 clusters at $z=0.54$ and 42 clusters at $z=0.82$. Because the mass function decreases as a power law with both the mass and the redshift of galaxy clusters, the limited volume of the \emph{MultiDark} simulation cube that enabled simulating the global properties of the MUSIC clusters is not large enough to populate the highest mass bin of the high redshift bin\footnote{This is also the case for the cluster sample of the NIKA2 SZ large program.} (see Fig. \ref{fig:MUSIC_NIKA2_sample}). In addition, the highest mass bin of the first redshift bin and the second-to-last mass bin of the high redshift bin cannot be completely filled for the same reason. Since the fraction of clusters with a disturbed ICM at high redshift is not a well known parameter, we chose to establish a sample containing nearly an equivalent number of dynamically relaxed and disturbed clusters to study the characteristics of the mean pressure profile of the two populations of clusters with an equivalent statistical power. A random selection would favor the selection of relaxed systems. Indeed, based on the three-dimensional morphological indicators presented in \cite{cia18}, the fractions of MUSIC synthetic clusters with a mass larger than $3\times 10^{14}~\mathrm{M_{\odot}}$ that are relaxed are equal to 63\% and 71\% at $z=0.54$ and $z=0.82,$ respectively. The other mass and redshift bins were thus populated by choosing two morphologically relaxed and two disturbed clusters among all MUSIC clusters available in each bin.\\
\begin{figure*}[h!]
\centering
\includegraphics[height=6.6cm]{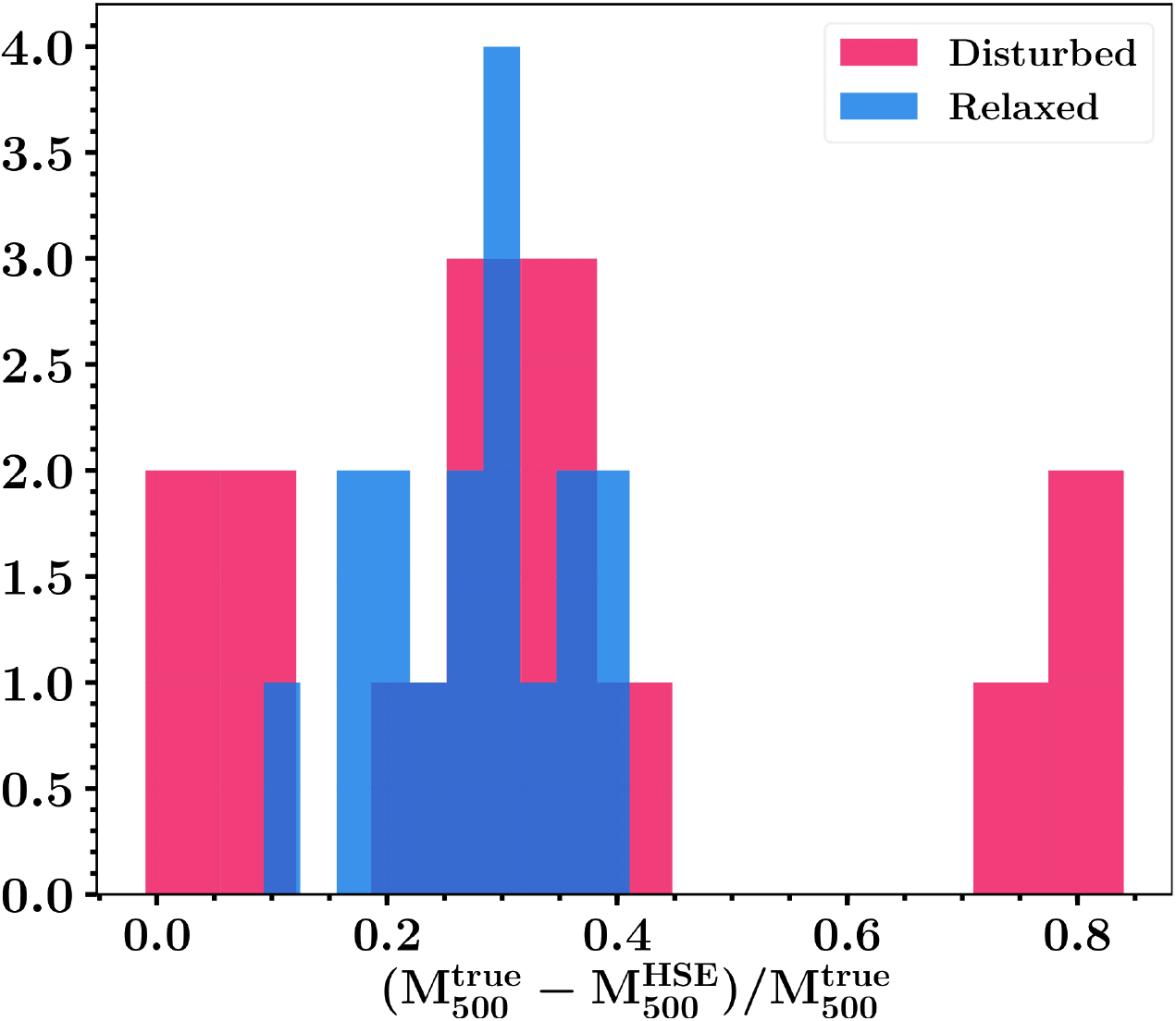}
\hspace{0.9cm}
\includegraphics[height=6.6cm]{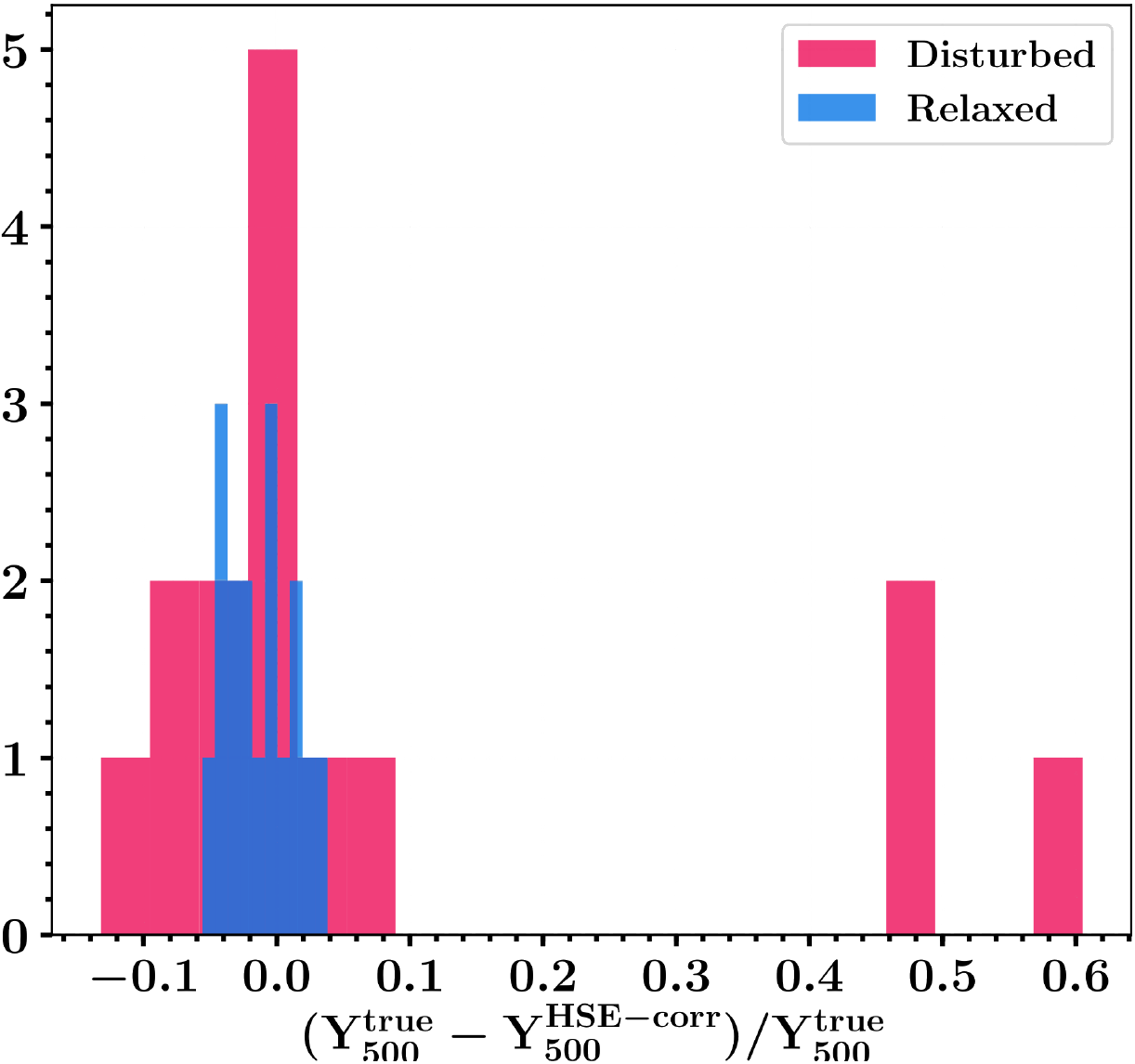}
\caption{{\footnotesize Distributions of the hydrostatic mass biases (left) and integrated Compton parameter biases (right) computed from the MUSIC profiles of the selected relaxed (blue) and disturbed (red) clusters. The methodology used to compute these quantities is explained in Sect. \ref{subsec:music_integ}.}}
\label{fig:MUSIC_NIKA2_biases}
\end{figure*}

Estimating the dynamical state of the MUSIC clusters by morphological considerations is first performed by a visual inspection of the Compton parameter maps in each bin. As shown on the left panel of Fig. \ref{fig:def_relax_disturbed}, a cluster is considered morphologically relaxed if it has a single tSZ surface brightness peak and its two-dimensional structure has a symmetry close to circular symmetry\footnote{This does not exclude halos with substructures aligned with the cluster center along the line of sight.}. On the other hand, the so-called disturbed clusters (see right panel of Fig. \ref{fig:def_relax_disturbed}) have a number of tSZ peaks greater or equal to two or an obvious deviation from circular symmetry. The $M$ morphological indicator introduced by \cite{cia18} to discriminate dynamically relaxed and disturbed clusters from Compton parameter maps is also used to confirm the selection made by visual inspection. This indicator combines several estimators of the dynamical state of the clusters (see \cite{cia18} for the full list of parameters). The $M$ indicator defined by the authors has been calibrated on the MUSIC sample after the segregation of the population for different dynamical states by common estimators applied in simulations. Dynamically relaxed clusters are such that $M<-0.41$ and disturbed clusters have $M>0.41$. The intermediate $M$ values form a mixed class in which the dynamical state of the clusters is not clearly identifiable.\\

We show the distributions of the $M$ indicator, discriminating the clusters between the relaxed and disturbed subsamples in the right panel of Fig. \ref{fig:MUSIC_NIKA2_sample}. The relaxed clusters in bin 1 have an average morphological indicator $M = -0.74 \pm 0.12$ and are indicated by blue dots in the left panel of Fig. \ref{fig:MUSIC_NIKA2_sample} and disturbed clusters (red triangles) of the same redshift bin are such that $M = 0.52 \pm 0.18$. The clusters of the second redshift bin are such that $M = -0.61 \pm 0.11$ for relaxed clusters and $M = 0.82\pm 0.31$ for disturbed clusters. We thus note an agreement between the visual characterization of the morphology of the selected clusters with the average results given by the $M$ indicator. As shown in the right panel of Fig. \ref{fig:MUSIC_NIKA2_sample}, the higher dispersion associated with the mean values of the $M$ indicator of the disturbed clusters is mainly caused by the two outliers at $M{\sim}2$ located in bin 1 and bin 2, respectively. These two clusters are ongoing mergers with a clear bimodal morphology and angular separations between the substructure SZ peaks enclosed between 2 and 3~arcmin. As it is also expected for the cluster sample of the NIKA2 SZ large program, some of the selected MUSIC clusters are defined by $M$ indicators in the intermediate morphology class highlighted by two vertical lines in the right panel of Fig. \ref{fig:MUSIC_NIKA2_sample}. We find four clusters defined as relaxed with $-0.41 < M < 0$ and seven clusters defined as disturbed with $0 < M < 0.41$. This higher fraction of disturbed clusters in the intermediate class is expected as there are only 33 MUSIC clusters at redshifts 0.54 and 0.82 with a mass larger than $3\times 10^{14}~\mathrm{M_{\odot}}$ that have $M > 0$ and 52\% of these have a morphology such that $M < 0.41$. We emphasize that our goal is not to define precise categories of cluster morphological states but to build a cluster sample that is not significantly biased toward a given morphology. Our methodology to define the relaxed and disturbed subsamples for the NIKA2 SZ large program may evolve and be slightly different from the method considered in this paper. The defined cluster sample finally has a fraction of relaxed clusters of about 50\%.\\
\begin{figure*}[h!]
\centering
\includegraphics[height=4.8cm]{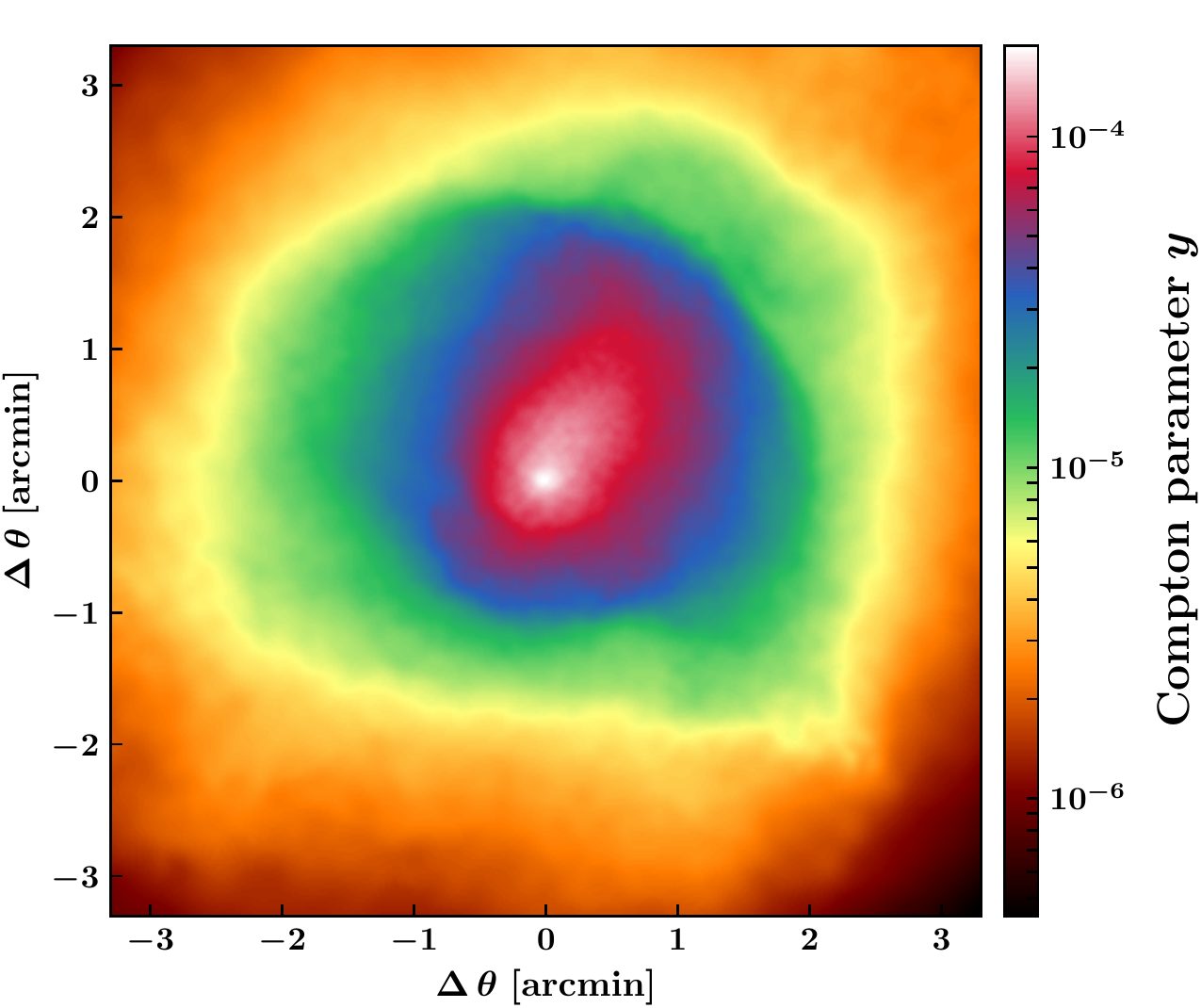}
\hspace{0.3cm}
\includegraphics[height=4.8cm]{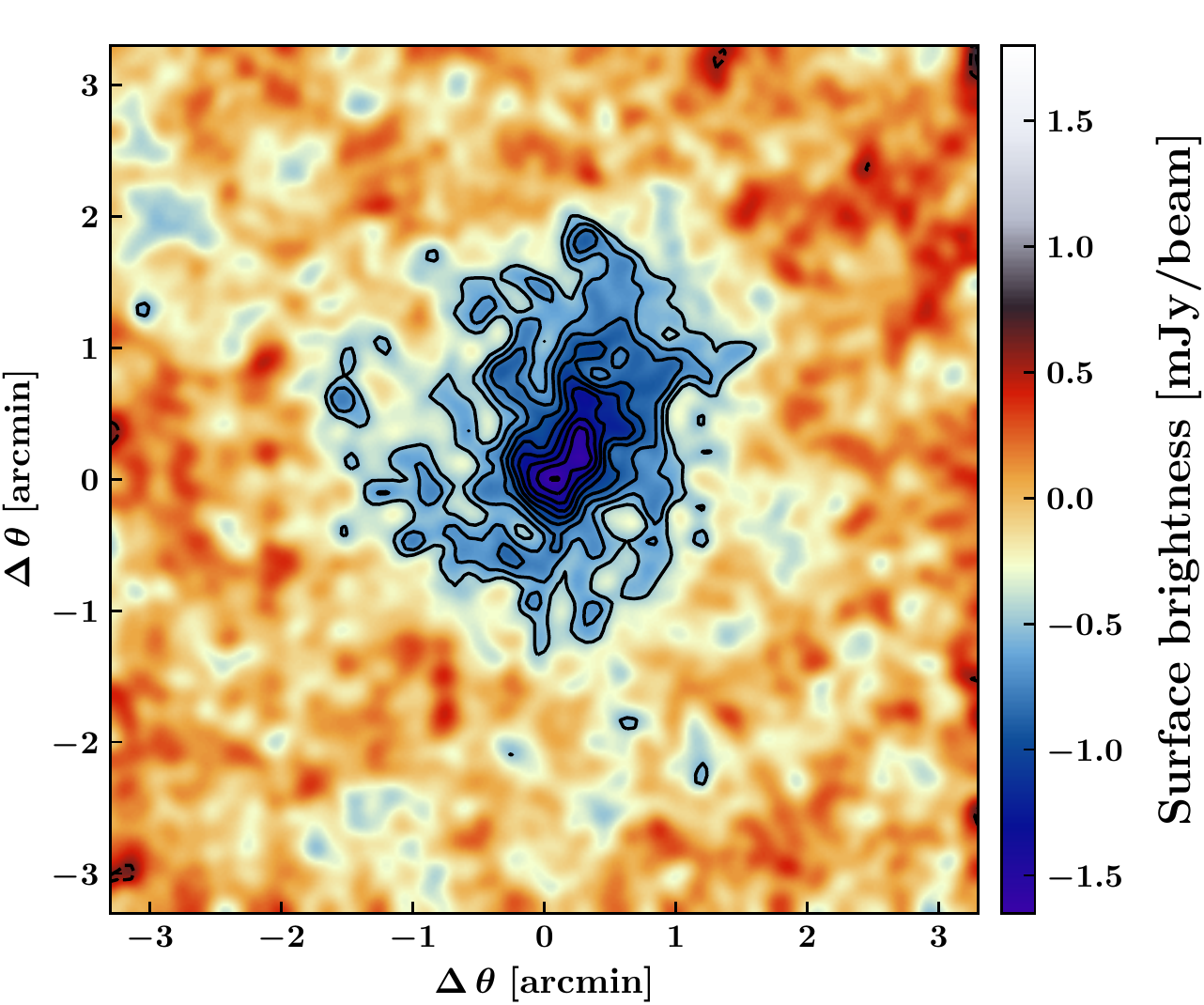}
\hspace{0.3cm}
\includegraphics[height=4.8cm]{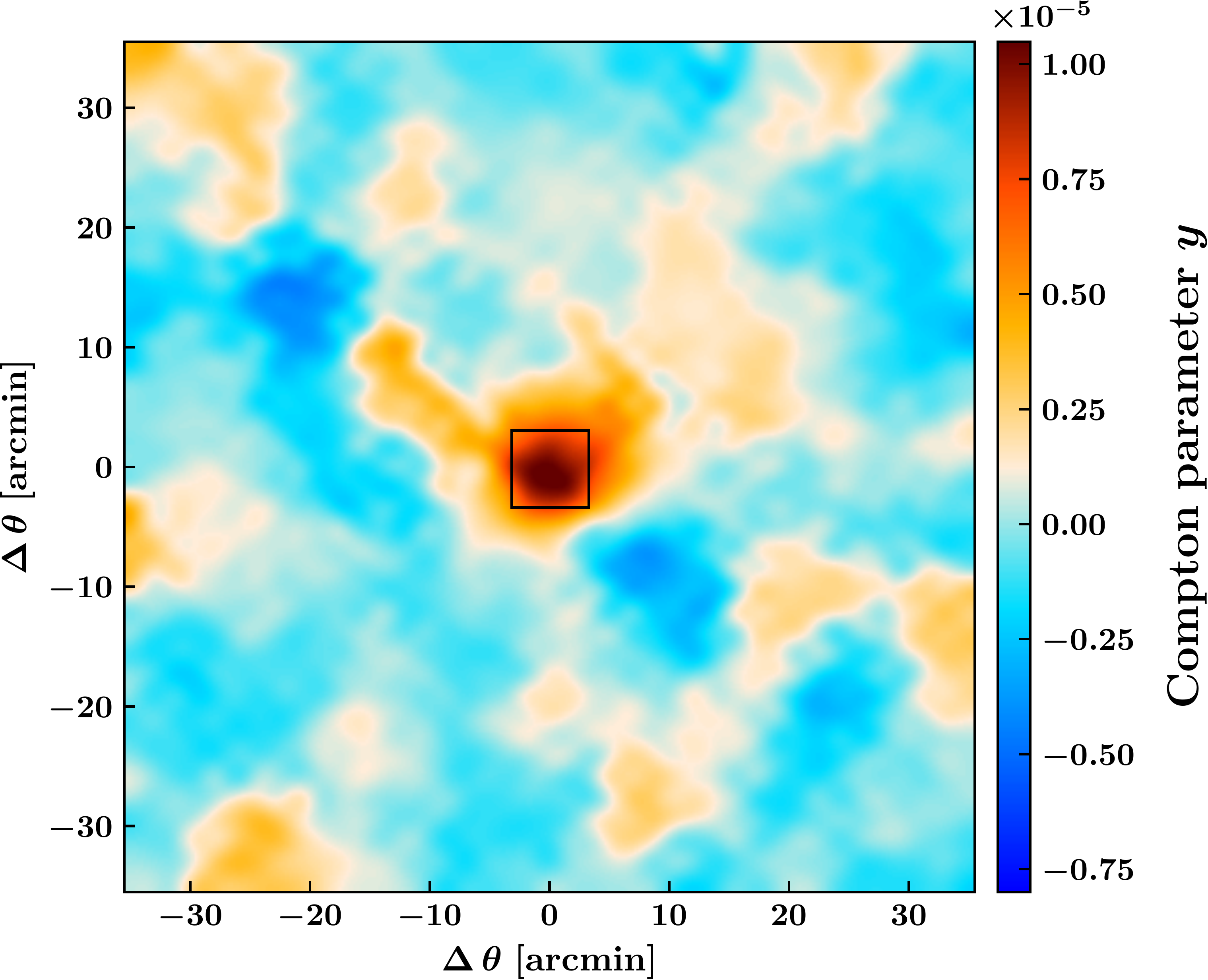}
\caption{{\footnotesize \textbf{Left:} MUSIC Compton parameter map of a selected disturbed cluster in the first redshift bin. An ICM extension is clearly identified in the  upper right region of the map.  \textbf{Middle:} Simulated NIKA2 tSZ surface brightness map at 150~GHz of the MUSIC cluster shown in the left panel. \textbf{Right:} Simulated \planck\ Compton parameter map of the MUSIC cluster shown in the left panel. The field of view considered for the left and middle panels is shown as a black rectangle at the center of the map.}}
\label{fig:simu_nk2_pla_maps}
\end{figure*}

The distributions of the biases on the mass and integrated Compton parameter of the relaxed and disturbed selected subsamples are shown in Fig. \ref{fig:MUSIC_NIKA2_biases}. These distributions were obtained following the method presented in Sect. \ref{subsec:music_integ} using the MUSIC profiles associated with the selected clusters. The median values of the hydrostatic bias and integrated Compton parameter are equal to 0.29 and -0.01 for both the distributions associated with the relaxed (blue) and disturbed (red) subsamples, respectively. Therefore, the hydrostatic bias and Compton parameter bias measured for the selected subsamples are compatible with the those measured on the distributions shown in Fig. \ref{fig:mass_Y500_bias} for the whole population of MUSIC clusters at $z=0.54$ and $z=0.82$. However, we note a larger dispersion of both the hydrostatic bias and integrated Compton parameter values for the disturbed clusters compared to the results obtained for the relaxed clusters. In particular, the two outliers with a morphological indicator $M{\sim}2$ (see Fig. \ref{fig:MUSIC_NIKA2_sample}) are also associated with high values of the hydrostatic and integrated Compton parameter biases. This is expected as the hydrostatic equilibrium hypothesis is far from being valid for these major mergers. Furthermore, we note the presence of a third outlier in the distributions shown in Fig. \ref{fig:MUSIC_NIKA2_biases}. This cluster has a morphological indicator $M=0.57$ and is therefore a member of the subsample of disturbed clusters. In addition, the ICM of this cluster is elongated to the southwest and the value of the Compton parameter is almost constant up to ${\sim}600$~kpc from the cluster core. The MUSIC density and pressure profiles of this cluster have a very shallow inner slope extending up to the same physical radius. The hydrostatic mass of this cluster is therefore significantly underestimated because the hydrostatic mass profile is very flat from the cluster center up to a large fraction of its virial radius. It is interesting to notice that some disturbed clusters present hydrostatic bias values close to 0. Shock regions around $R_{500}$ are usually found in such clusters and are responsible for important discontinuities in their pressure profile. This results in a significant increase of the hydrostatic mass profile around $R_{500}$ which then stabilizes rapidly at larger radii. For this reason, it is possible to measure values of $M_{500}^{\mathrm{HSE}}$ that are very close\footnote{They can also be larger as shown in the left panel of Fig. \ref{fig:mass_Y500_bias}.} to the true mass $M_{500}^{\mathrm{true}}$, although a higher value of the hydrostatic bias would be measured at radii larger that $R_{500}$ in these systems. While the scatter associated with the distribution of the integrated Compton parameter biases for our subsample of disturbed clusters is seven times larger than the scatter measured for the relaxed clusters, we note that it depends significantly on the dynamical state of the selected clusters. Indeed, if we discard the three outliers in the distribution shown in the right panel of Fig. \ref{fig:MUSIC_NIKA2_biases}, the dispersion of the biases on the integrated Compton parameters of the disturbed clusters around the median is only twice larger than that measured for the relaxed clusters.\\

The selected sample of MUSIC clusters is processed in Sect. \ref{sec:mean_prof} as the sample of clusters from the NIKA2 SZ large program would be in order to constrain the mean ICM pressure profile for both the relaxed and disturbed subsamples.

\subsection{Simulation of NIKA2 and \planck\ tSZ observations}\label{subsec:simu_obs}

This section describes the procedure used to compute the NIKA2 tSZ surface brightness and \planck\ Compton parameter maps from the Compton parameter maps associated with the selected MUSIC clusters. As we want to focus this analysis on the impact of the dynamics of the selected clusters on the mean pressure profile, we do not consider the contamination of the tSZ signal induced by the radio or submillimeter emission of point sources. These contaminants however increase the uncertainties associated with the pressure profiles deprojected from the NIKA2 and \planck\ maps of the clusters observed for the NIKA2 SZ large program \citep[{e.g.,}][]{ada16a}.\\

The MUSIC Compton parameter maps are first converted into tSZ surface brightness maps by applying the conversion coefficient given by integrating the spectrum of the tSZ effect into the NIKA2 bandpass at 150~GHz. The spatial distribution of the tSZ signal in the resulting maps is then convolved by the 17.7~arcsec FWHM Gaussian beam and the NIKA2 transfer function at 150~GHz obtained from the first NIKA2 tSZ observations \citep{rup18} to take into account the different filtering effects induced by the observations and the raw data analysis. The typical observation time considered for the clusters in each mass and redshift bin of the NIKA2 SZ large program is used to produce a map of the expected standard deviation per pixel associated with each MUSIC selected cluster. These observation times are estimated based on the respective integrated Compton parameter values $Y_{500}^{\mathrm{HSE}}$ of each cluster and the NIKA2 sensitivity at 150~GHz. These standard deviation maps and the noise power spectrum of the residual correlated noise observed in typical NIKA2 tSZ maps \citep{rup18} allow us to simulate residual noise maps for each selected cluster. The sum of the filtered tSZ surface brightness signal and residual noise allows for the production of realistic NIKA2 tSZ maps of the selected MUSIC clusters. We note that CMB has a negligible contribution at the angular scales that can be recovered by NIKA2. Furthermore, as shown in \cite{rup18}, the cosmic infrared background (CIB) signal is an order of magnitude lower than the RMS noise otherwise caused by both the instrumental and atmospheric noise contributions. Therefore, we chose to ignore these contaminants in the simulation of the NIKA2 maps in this work. A potential improvement of the simulated observations would consist of including radio and submillimeter sources in the field. We did not include point source contaminants in our simulations in order to focus our study on the impact of cluster dynamics on the mean pressure profile.\\

The \planck\ Compton parameter maps of the selected MUSIC clusters are obtained with a similar procedure. The distribution of the tSZ signal in the MUSIC maps is first convolved by the 10~arcmin FWHM \planck\ beam. The pixels of the resulting maps are then combined to form 1.7~arcmin large pixels to limit the size of the simulated \planck\ maps while maintaining a sufficient number of pixels per beam. We estimated \planck\ noise maps for each cluster by considering the \planck\ noise power spectrum and different sky coordinates randomly drawn in the area observable by NIKA2, {i.e.,} $\mathrm{dec}> -11^{\circ}$. This allowed us to take into account both the spatial correlation and the variations in the amplitude of the residual noise in the \planck\ $y$-map given the  position considered on the sky. The sum of the maps of the tSZ signal, smoothed at the \planck\ angular resolution, and of the correlated noise results in realistic \planck\ Compton parameter maps of the selected MUSIC clusters. We note that the tSZ signal outside the MUSIC field of view of about 23~arcmin is set to zero in this procedure. However the tSZ signal at a projected radius of 10~arcmin from the cluster center is already negligible in comparison to the \planck\ RMS noise at this distance. Therefore, we did consider all the relevant tSZ signal in the simulation of the \planck\ maps.\\

The comparison of the simulated NIKA2 and \planck\ tSZ maps in Fig. \ref{fig:simu_nk2_pla_maps} highlights the complementarity of these two experiments regarding the characterization of the spatial distribution of the tSZ signal of high redshift clusters. The left panel of Fig. \ref{fig:simu_nk2_pla_maps} shows the Compton parameter map of a MUSIC cluster with a mass $M_{500} = 5.5 \times 10^{14}~\mathrm{M_{\odot}}$ at redshift $z = 0.54$ considering a field of view of 6.5~arcmin. As shown in the middle panel of the figure, the high angular resolution of the NIKA2 camera enables us to map the tSZ signal of the cluster up to a projected distance of about 2~arcmin from the tSZ peak. The large scale structures of the tSZ signal spatial distribution are lost because of the important filtering induced by NIKA2 on these scales. They are however partially recovered in the \planck\ Compton map (see right panel of Fig. \ref{fig:simu_nk2_pla_maps}), which does not provide any information on the internal structure of the ICM but enables the anchoring of the total integrated Compton parameter of the cluster. The simulated NIKA2 tSZ surface brightness and \planck\ Compton parameter maps constitute the data set used in the analysis developed in Sect. \ref{sec:mean_prof}.

\section{Characterization of the mean pressure profile of the synthetic sample}\label{sec:mean_prof}

This section presents the analysis procedure and results obtained concerning the estimation of the mean pressure profile of the selected MUSIC clusters from the simulated NIKA2 and \planck\ tSZ maps. The impact of ICM disturbances on the individual profiles and the fraction of disturbed clusters on the mean pressure profile is discussed and placed within the framework of the future results of the NIKA2 SZ large program.

\subsection{Estimation of the individual pressure profiles from the simulated tSZ maps}\label{subsec:profile_extract}

\begin{figure*}[h!]
\centering
\includegraphics[height=6.6cm]{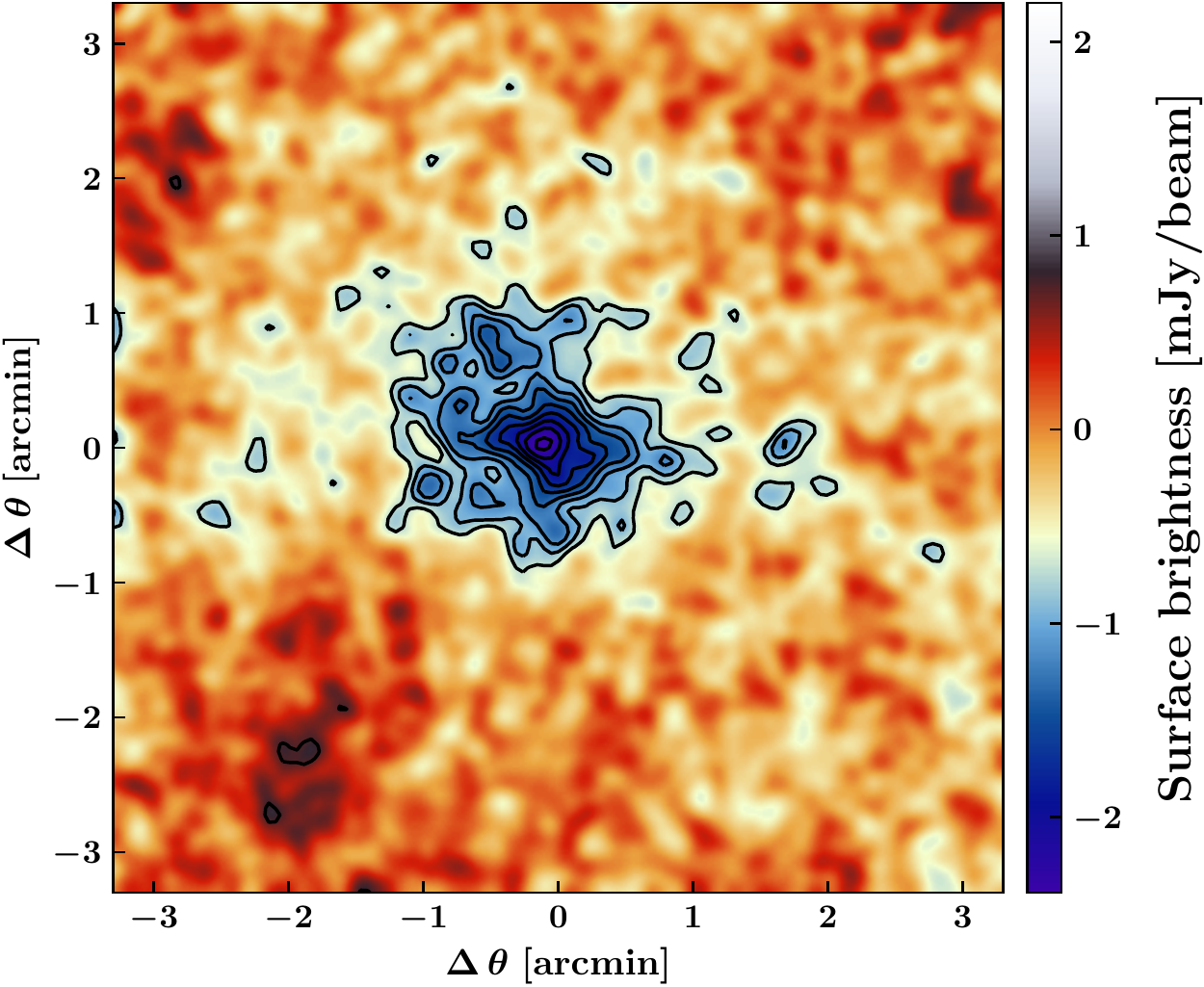}
\hspace{1.2cm}
\includegraphics[height=6.6cm]{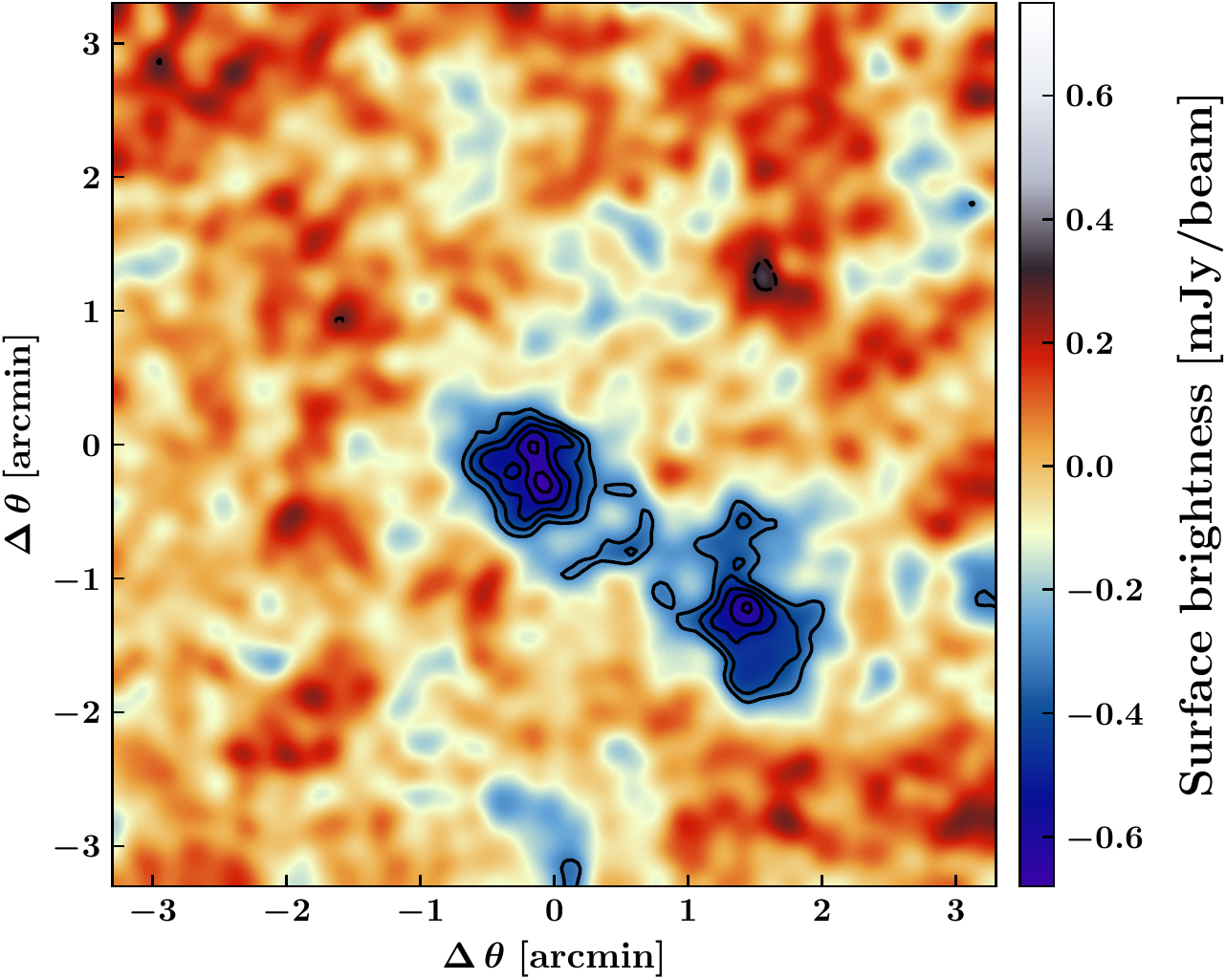}
\includegraphics[height=6.6cm]{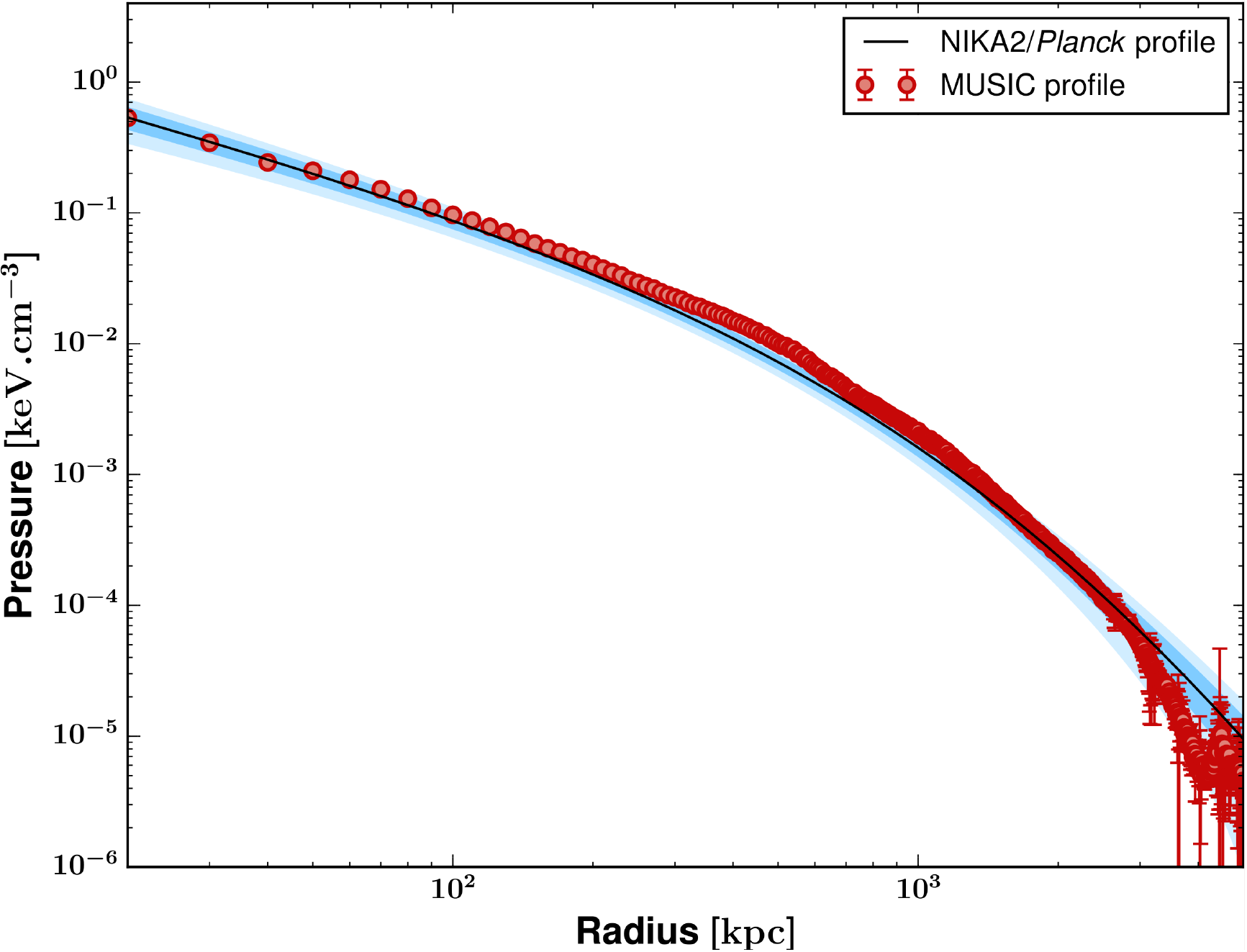}
\hspace{0.6cm}
\includegraphics[height=6.6cm]{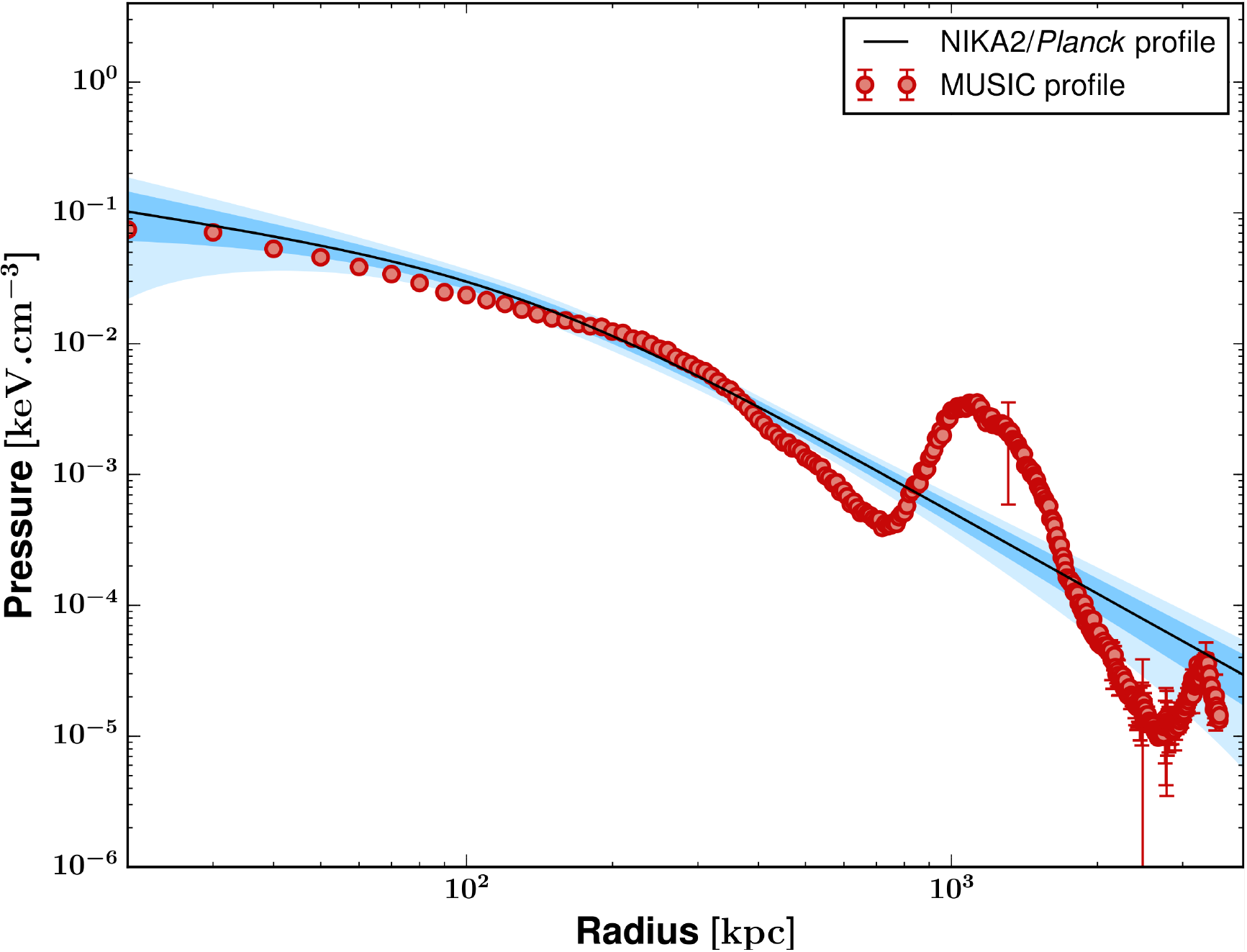}
\caption{{\footnotesize \textbf{Upper panels:} Simulated NIKA2 tSZ surface brightness maps for a relaxed (left) and a disturbed (right) cluster. The signal-to-noise contours (black lines) start at $3\sigma$ with $1\sigma$ steps. \textbf{Lower panels:} Pressure profiles estimated at the maximum likelihood from the MCMC analysis of the maps shown in the above panels (black line) and associated 1 and $2\sigma$ uncertainties (dark blue and light blue regions). The pressure profiles extracted from the MUSIC simulation for each cluster are represented by the red dots with $1\sigma$ error bars. A radius of 100~kpc corresponds to an angular scale of 15.3~arcsec and 12.8~arcsec for the left and right panel, respectively.}}
\label{fig:results_relax_disturbed}
\end{figure*}

The simulated NIKA2 and \planck\ tSZ maps are jointly analyzed with the NIKA2 tSZ analysis pipeline \citep{rup18b} based on a Markov chain Monte Carlo (MCMC) procedure using the Python package emcee \citep{for13} as described in detail in \citep{rup18}. A review of the main steps of the analysis is provided in this section. The pressure distribution within the ICM of each selected MUSIC cluster is modeled by a gNFW pressure profile given the results obtained in Sect. \ref{subsec:music_prof}. At each iteration of the MCMC algorithm, the pressure distribution is integrated along the line of sight to compute a model of the Compton parameter profile of the considered cluster (see eq. \ref{eq:y_compton}). The Compton parameter profile is then convolved with the beam profile of NIKA2 modeled as a two-dimensional Gaussian function with a 17.7~arcsec FWHM. The convolved profile is then projected on the pixelized grid considered for the NIKA2 simulated data. We applied the NIKA2 transfer function to the computed Compton parameter map to account for the processing filtering that suppresses signal on large scales. The filtered Compton parameter map is converted into a NIKA2 tSZ surface brightness map using a conversion coefficient obtained by integrating the tSZ spectrum within the 150 GHz NIKA2 bandpass. This conversion coefficient is associated with a flux calibration uncertainty of 10\% and varies within a Gaussian prior in the analysis. \\
As almost all the scales of the spatial distribution of the tSZ signal are smoothed by the \planck\ beam, we used the integrated Compton parameter $Y_{\mathrm{5R500}}$, estimated by aperture photometry on the simulated \planck\ maps, to average the information contained in all the pixels of each \planck\ map in a single measurement point. The model of the integrated Compton parameter is given by the spherical integral of the current pressure profile up to $5R_{500}$. We then compared the NIKA2 tSZ map model $\tilde{M}$ and the \planck\ integrated Compton parameter model $\tilde{Y}$  with the NIKA2 mock data $M_{\mathrm{NIKA2}}$ and the integrated Compton parameter measured on the simulated \planck\ map using the following likelihood function:
\begin{equation}
\begin{tabular}{rl}
        $-2 \mathrm{ln} \, \mathscr{L}$  & $ =\chi^2_{\mathrm{NIKA2}} + \chi^2_{\mathrm{Planck}}$\\[0.2cm]
         &$ =\sum_{i=1}^{N_{\mathrm{pixels}}^{\mathrm{NIKA2}}} [(M_{\mathrm{NIKA2}} - \tilde{M})^T C_{\mathrm{NIKA2}}^{-1} (M_{\mathrm{NIKA2}} - \tilde{M})]_i $ \\[0.2cm]
         & $ + \left[\frac{Y_{\mathrm{5R500}} - \tilde{Y}}{\Delta Y_{\mathrm{5R500}}}\right]^2 $
\label{eq:chi2_NK2_Planck_MUSIC}
\end{tabular}
,\end{equation}
where the uncertainty $\Delta Y_{\mathrm{5R500}}$ associated with the $Y_{\mathrm{5R500}}$ data point measured on the simulated \planck\ map corresponds to the dispersion of aperture photometry measurements performed around the cluster position where the noise is homogeneous. We stress that the flux calibration uncertainty is included in this likelihood as a Gaussian prior on the conversion coefficient used to compute $M_{\mathrm{NIKA2}}$. The NIKA2 correlated noise power spectrum considered to simulate the tSZ maps is used to estimate the noise covariance matrix $C_{\mathrm{NIKA2}}$ in the likelihood function (\ref{eq:chi2_NK2_Planck_MUSIC}). The MCMC analysis performed for each cluster is made by considering a total of 200 Markov chains on 20 CPUs in parallel. The convergence test of \cite{gel92} is used to stop the MCMC sampling of the parameter space. Although the authors of emcee do not recommend using this test, we checked that the final chains also present short autocorrelation time values. The computation time required for the MCMC analysis to be completed is on the order of 2 to 3 days per cluster. At the end of each analysis the pressure profile obtained at the maximum likelihood is stored in a file along with the associated $1\sigma$ uncertainty on the pressure profile. The uncertainties are estimated by Monte Carlo sampling of the posterior using the Markov chains associated with each parameter of the gNFW model.\\

The maximum-likelihood pressure profile is compared with the pressure profile extracted directly from the MUSIC simulation (see Sect. \ref{subsec:music_prof}). The upper and lower left panels in Fig. \ref{fig:results_relax_disturbed} present the simulated NIKA2 tSZ surface brightness map for a relaxed cluster of the penultimate mass bin of the low redshift bin and its associated pressure profile, respectively. The black curve corresponds to the pressure profile obtained at the maximum likelihood and the uncertainties at 1 and $2\sigma$ are given by the dark and light blue regions. The red dots correspond to the pressure profile extracted from the MUSIC simulation for this cluster. The profile constrained by the MCMC analysis using a deprojection of the tSZ signal contained in the NIKA2 and \planck\ simulated maps is therefore compatible with the radial pressure distribution of the considered MUSIC cluster. The slight relative differences observed at certain radii are due to deviations of the shape of the MUSIC pressure profile with respect to the smooth pressure distribution given by the gNFW model. These results validate the procedure used in the NIKA2 tSZ analysis pipeline to estimate the pressure distribution inside galaxy clusters from the combination of complementary tSZ data sets.\\

The right panels in Fig. \ref{fig:results_relax_disturbed} correspond to the tSZ surface brightness map of a morphologically disturbed cluster of the second mass bin in the second redshift bin (top) and its associated pressure profile (bottom). Since this cluster is clearly bimodal and in a merger state in the NIKA2 simulated map, a single gNFW model does not allow us to constrain all the features of the ICM radial pressure distribution extracted from the MUSIC simulation. As shown by the comparison between the profile obtained at the maximum likelihood and the profile extracted from the MUSIC simulation, the pressure distribution estimated from the MCMC analysis is significantly different from the mean radial distribution obtained in spherical shells for radii between 700 and 1500~kpc. Although the use of a nonparametric profile would enable us to constrain part of this deviation from the decreasing shape of the gNFW model, it is important to note that any model based on a radial pressure distribution is not suitable for this particular cluster because the circular symmetry of the spatial distribution of the tSZ signal is broken. A way to characterize the bias induced by this mismodeling of the pressure distribution would be to mask, in turn, the identified  main substructures of the disturbed clusters and perform the same analysis for each of these in order to compute an additional systematic uncertainty on the estimated pressure profile; see \cite{rup18}. This type of analysis will be carried out for the clusters of the NIKA2 SZ large program but is beyond the scope of this study as all the clusters defined as a single halo in the simulation have to be modeled by a single pressure profile to preserve consistency in a cosmological analysis (see Sect. \ref{subsec:goal_szlp}). We also note that the pressure profile estimated from the simulated tSZ observations of this cluster leads to a relative difference on the measurement of $\hat{Y}_{500}^{\mathrm{HSE}}$ of 67\% compared to the expected value obtained by considering the \cite{arn10} universal pressure profile and the measurement of $\hat{Y}_{\mathrm{5R500}}$ computed by aperture photometry on the simulated \planck\ map. This result highlights the impact of the NIKA2 high angular resolution on the estimation of the integrated parameter of high redshift clusters.\\

At the end of the MCMC analysis of each simulated tSZ map, the constrained pressure profile is combined with the density profile extracted from the MUSIC simulation\footnote{We treat the MUSIC density profile as a profile that would be estimated from X-ray observations.} to estimate a mass profile under the hydrostatic equilibrium hypothesis using eq. (\ref{eq:mass_HSE}). This mass profile is used to compute the value of the characteristic radius $\hat{R}_{500}^{\mathrm{HSE}}$ considered to estimate the integrated quantities $\hat{M}_{500}^{\mathrm{HSE}}$, $\hat{Y}_{500}^{\mathrm{HSE}}$, and $\hat{P}_{500}^{\mathrm{HSE}}$ (see Sect. \ref{subsec:music_integ}). These quantities are essential to estimate the mean pressure profile from all pressure profiles obtained by this analysis.

\begin{figure*}[h!]
\centering
\includegraphics[height=6.4cm]{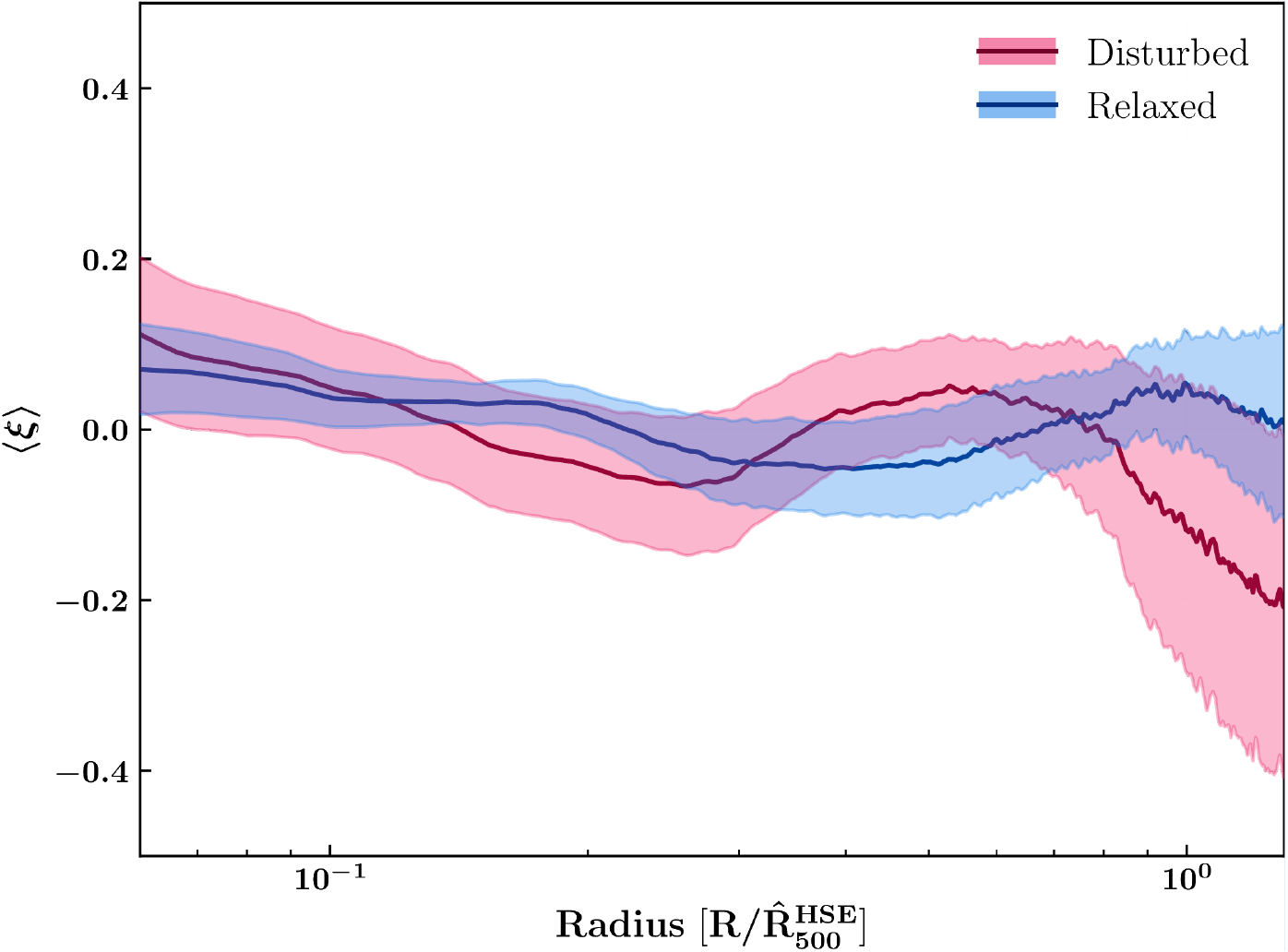}
\hspace{0.6cm}
\includegraphics[height=6.4cm]{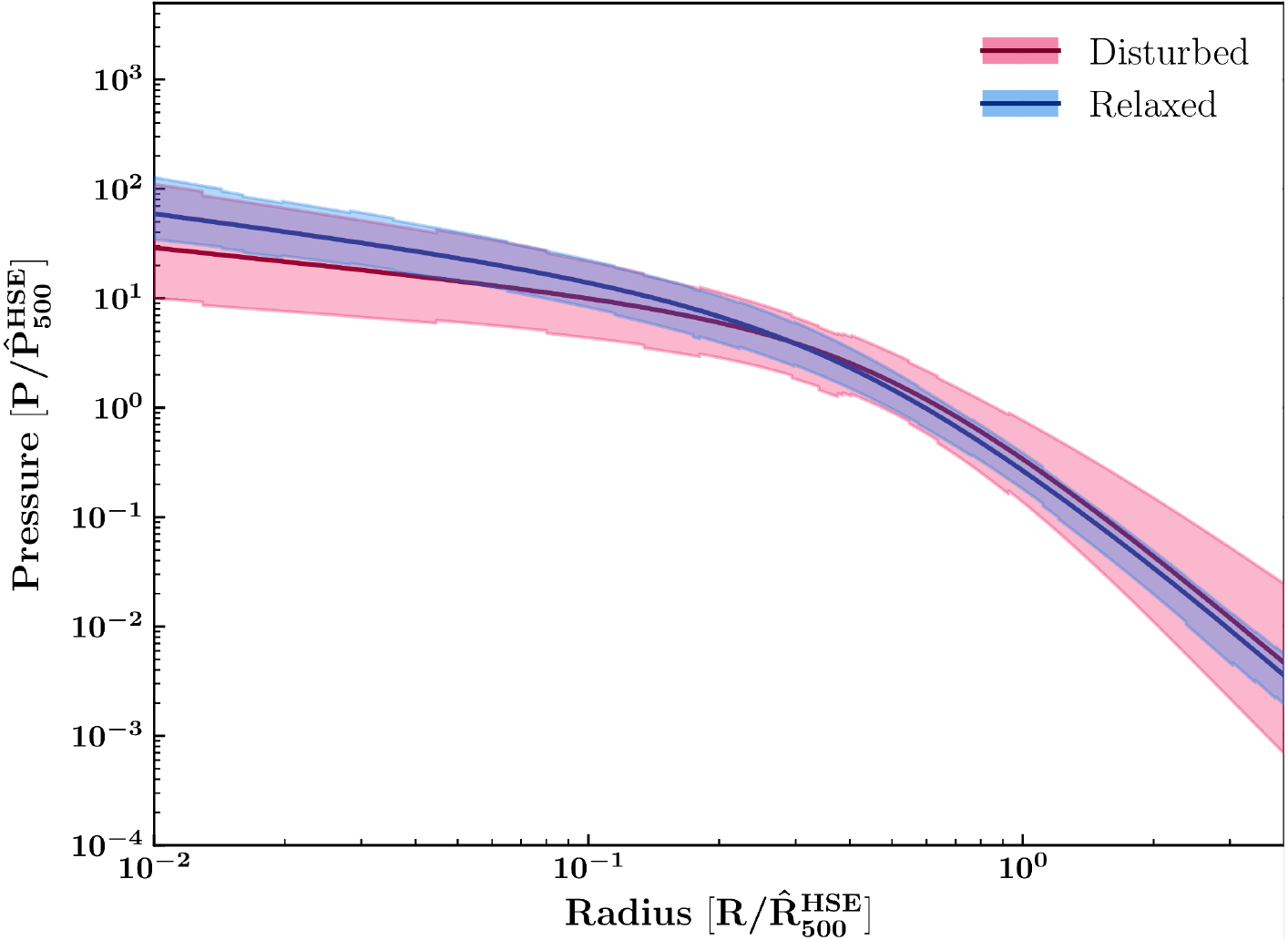}
\caption{{\footnotesize \textbf{Left:} Mean relative difference between the NIKA2/\planck\ deprojected pressure profiles and those extracted from the MUSIC simulation, presented as a function of the normalized radius for the relaxed (blue) and the disturbed (red) cluster subsamples. The shaded regions represent the $1\sigma$ error on the mean. \textbf{Right:} Mean normalized pressure profiles and associated $1\sigma$ scatter obtained from the profile distributions of relaxed (blue) and disturbed (red) clusters.}}
\label{fig:mean_pressure_prof}
\end{figure*}

\subsection{Impact of the ICM dynamical state on the mean pressure profile}\label{subsec:impact_icm_dist}

This section is dedicated to the analysis of the impact of the ICM dynamical state  of the selected MUSIC clusters on their pressure profiles estimated from the analysis developed in Sect. \ref{subsec:profile_extract} and on the mean pressure profile of the relaxed and disturbed cluster subsamples.\\
The pressure profiles, estimated at the end of the analysis described in Sect. \ref{subsec:profile_extract}, are normalized using the values of $\hat{R}_{500}^{\mathrm{HSE}}$ and $\hat{P}_{500}^{\mathrm{HSE}}$ computed for each cluster by the combination of the MUSIC density profile and the NIKA2/\planck\ deprojected pressure profile. As the uncertainties associated with the estimated pressure profiles for the disturbed clusters do not include the large modeling errors induced by the strong deviations from spherical symmetry and hydrostatic equilibrium (see Sect. \ref{subsec:profile_extract}), we decided to compare the estimated pressure profiles $\tilde{P}$ with the MUSIC profiles extracted from the simulation $P_{\mathrm{MUSIC}}$ by computing the relative difference, i.e.,
\begin{equation}
\xi = \frac{P_{\mathrm{MUSIC}} - \tilde{P}}{P_{\mathrm{MUSIC}}}
.\end{equation}
The variations of $\xi$ as a function of $R/\hat{R}_{500}^{\mathrm{HSE}}$ for all the selected MUSIC clusters are then used to compute the mean and the error on the mean of the distributions of $\xi$ measured at each scaled radius for the disturbed and the relaxed clusters. The left panel of Fig. \ref{fig:mean_pressure_prof} shows the variations of the mean value of $\xi$ as a function of the normalized radius $R/\hat{R}_{500}^{\mathrm{HSE}}$ for the relaxed (blue) and disturbed (red) simulated clusters. The shaded blue and red regions give the $1\sigma$ errors on the mean values of $\xi$ computed at each radius as the standard deviation of the distributions divided by the square root of the number of relaxed and disturbed profiles, respectively. The relative difference between the deprojected pressure profile and that extracted in the three-dimensional volume of the simulation is always lower than 10\% in the radius range where the tSZ signal is not filtered significantly by NIKA2, {i.e.,} for $R$ between $0.1 \, \hat{R}_{500}^{\mathrm{HSE}}$ and $\hat{R}_{500}^{\mathrm{HSE}}$. Furthermore, the average of the mean relative difference in this radius range is equal to 0.9\% and 0.05\% for the disturbed and the relaxed clusters respectively. The variations of $\xi$ around 0 are due to deviations of the shape of the simple gNFW model from the true pressure distribution of the simulated clusters. Although these deviations tend to be small for relaxed clusters (see the left panels of Fig. \ref{fig:results_relax_disturbed}), they can reach an order of magnitude at specific radii for disturbed systems (see the right panels of Fig. \ref{fig:results_relax_disturbed}). This explains why the error on the mean of $\xi$ associated with the disturbed clusters is on average 90\% larger than that measured for the relaxed clusters. We note however that the mean relative difference computed for the disturbed clusters in the radius range of interest for NIKA2 is not significantly larger than that measured for the relaxed clusters. This comes from the fact that the location of unvirialized structures in the ICM of the disturbed clusters is not always the same. The large relative differences measured in a given radius range thus tend to compensate each other for a sample of disturbed systems. This implies that the mean normalized pressure profiles obtained by analyzing the simulated NIKA2 and \planck\ data for the two subsamples are compatible with the mean profiles obtained by combining the profiles extracted directly from the MUSIC simulation for the selected clusters.\\
\begin{figure*}[h!]
\centering
\includegraphics[height=6.4cm]{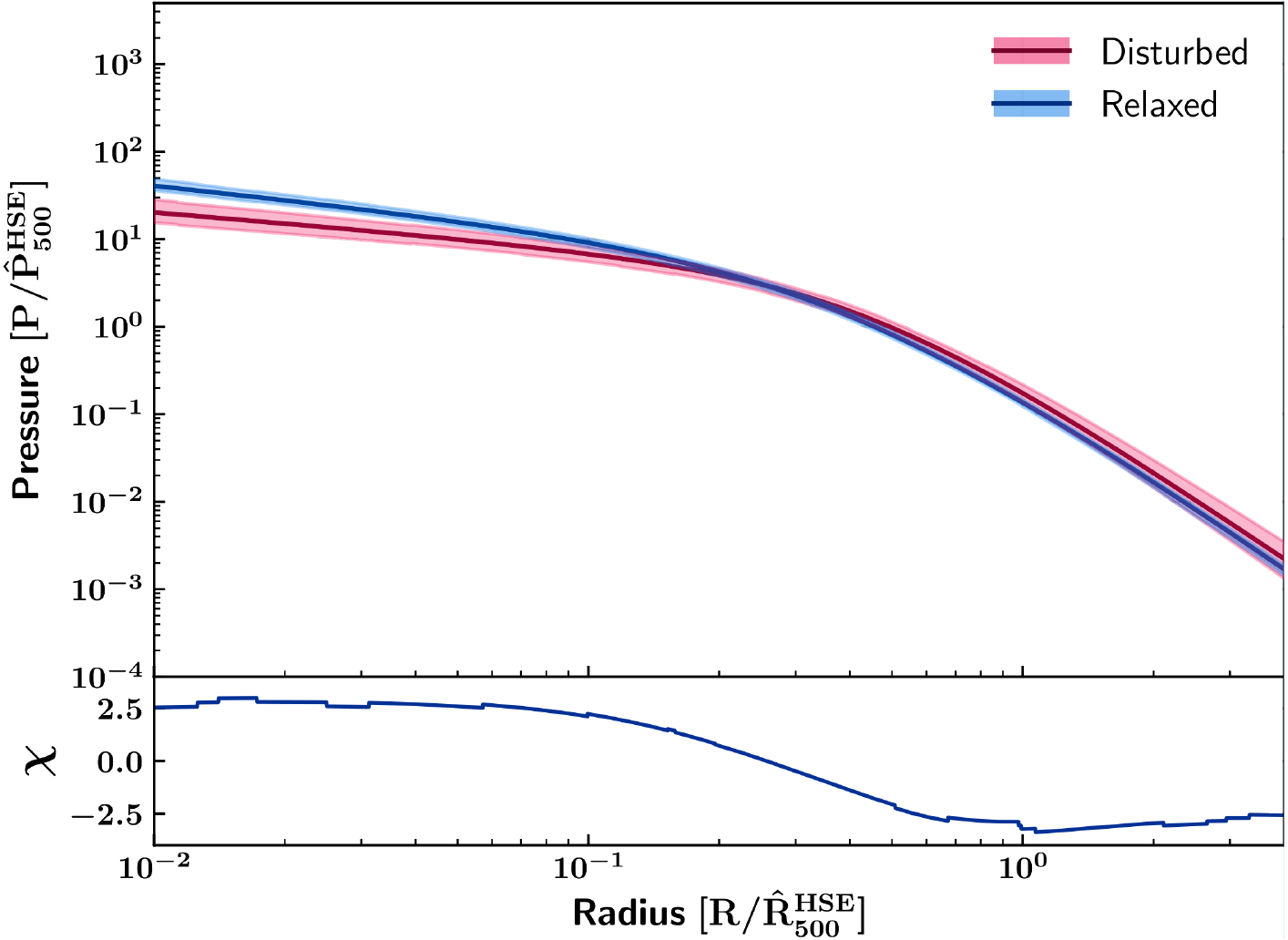}
\hspace{0.6cm}
\includegraphics[height=6.4cm]{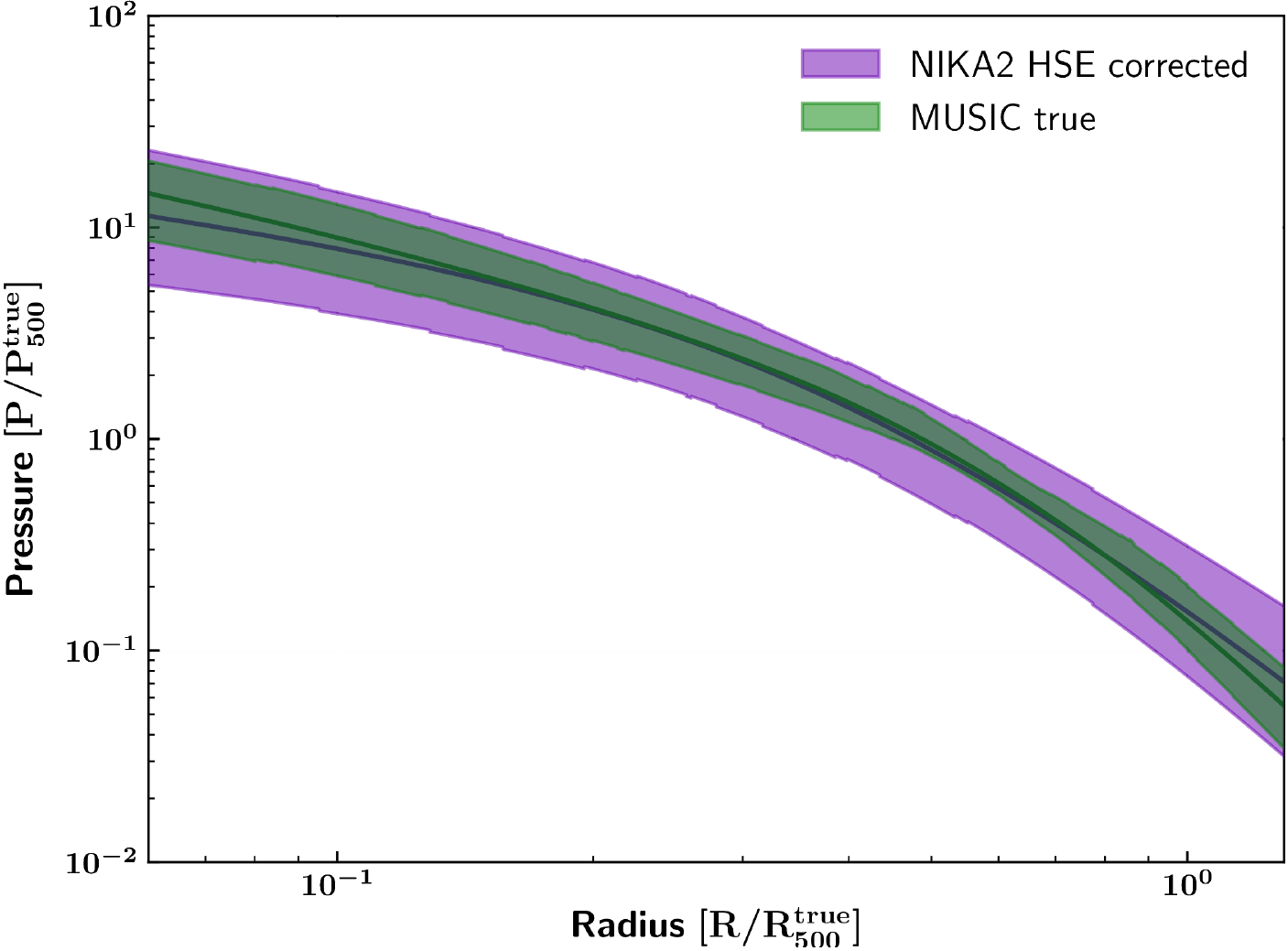}
\caption{{\footnotesize \textbf{Left:} Mean normalized pressure profiles and associated $1\sigma$ error on the mean for the relaxed (blue) and disturbed (red) clusters. The difference between the two profiles divided by the mean of their associated uncertainties is shown with the dark blue line in the lower panel. \textbf{Right:} Mean normalized pressure profiles and associated $1\sigma$ scatter for the whole sample of selected clusters computed from the true MUSIC pressure profiles (green) and for the NIKA2/\planck\ deprojected profiles (purple) after normalization by the integrated quantities corrected from the mean hydrostatic bias (see Sect. \ref{subsec:music_integ}).}}
\label{fig:final_compare}
\end{figure*}

The distribution of the normalized pressure profiles for the relaxed and disturbed selected clusters are treated separately to estimate the mean pressure profile associated with the two populations at high redshift using the method developed in the Sect. \ref{subsec:music_prof}. The pressure values in each profile are weighted by the inverse of their associated error bars in the calculation of the mean profile (see Sect. \ref{subsec:profile_extract}). The resulting mean profiles are represented with the blue and red lines in the right panel of Fig. \ref{fig:mean_pressure_prof}. The blue and red shaded regions represent the scatter of the distributions of normalized pressure profiles for the relaxed and disturbed clusters, respectively. The flattening of the mean pressure profile associated with the disturbed clusters in the core of the ICM compared to the profile associated with the relaxed clusters is consistent with the trend observed in different observational studies \citep[{e.g.,}][]{arn10}. The scatter associated with the mean profiles in the central region of the clusters ($r < 0.1\mathrm{R_{500}}$) and their outskirts ($r > 2\mathrm{R_{500}}$) are about 20 and 30\% higher than those associated with the mean profiles estimated from the MUSIC profiles normalized by the same integrated quantities for the relaxed and disturbed clusters, respectively. The increase of the scatter observed for the relaxed clusters is due to the angular resolution of NIKA2 at 150~GHz and its 6.5~arcmin field of view, which do not allow the central parts of high redshift clusters to be resolved and their outskirts to be mapped without significant filtering. For disturbed clusters, the observed increase of the scatter is more significant, especially in the outskirts, because the constraints on the external slopes of the individual profiles are affected by the presence of substructures and deviations from circular symmetry. The scatter associated with the mean profile of the disturbed clusters is 65\% greater than that observed for the profile distribution of the relaxed clusters at a radius $R = \mathrm{R_{500}}$. This result shows that a variation of the fraction of disturbed clusters with redshift can induce a significant change of the measured intrinsic scatter associated with the distribution of pressure profiles.\\

We present in the left panel of Fig. \ref{fig:final_compare} the mean normalized pressure profiles of the relaxed and disturbed subsamples and the associated error on the mean. This error is computed from the ratio of the intrinsic scatter of each distribution and the square root of their respective number of profiles. The lower panel of the figure shows the difference between the two mean pressure profiles divided by the average of the errors on the means at each radius. For this particular sample, for which the fraction of disturbed clusters is about 50\%, we show that differences in the mean pressure profile of relaxed and disturbed clusters can be identified by NIKA2 at a significance higher than $2\sigma$ for most of the scales recovered by the camera. This significance would have been higher had our selected sample been built using the same statistics of the NIKA2 SZ large program sample, {i.e.,} five clusters per bin instead of four (see Sect. \ref{subsec:nika2_sample}). Furthermore, given the results on the intrinsic scatter associated with the mean pressure profiles of the relaxed and disturbed subsamples (see Fig. \ref{fig:mean_pressure_prof}), we expect this significance on the difference between the mean pressure profiles to increase if the actual fraction of disturbed clusters in the cluster sample of the NIKA2 SZ large program is between 50\% and 70\%.\\

Finally, we compared the mean pressure profile of the whole selected cluster sample after a normalization by the corrected integrated quantities $\hat{R}_{500}^{\mathrm{HSE-corr}}$ and $\hat{P}_{500}^{\mathrm{HSE-corr}}$ with the true mean normalized pressure profile computed from the MUSIC pressure profiles and the true integrated quantities $R_{500}^{\mathrm{true}}$ and $P_{500}^{\mathrm{true}}$ of each cluster in the selected sample. The results are shown in the right panel of Fig. \ref{fig:final_compare} with the purple and green profiles, respectively. As expected given the previous results, we do not measure significant differences between the mean profiles in the scaled radius interval that can be probed by the NIKA2 SZ large program {i.e.,} $0.1\mathrm{R_{500}} < r < \mathrm{R_{500}}$. However, the intrinsic scatter associated with the NIKA2/\planck\ deprojected profiles (purple shaded region) is on average twice as large as that measured on the true distribution of normalized profiles (green shaded region). While 60\% of this increase at $R_{500}$ can be explained by the normalization of the profiles using integrated quantities corrected by the mean hydrostatic bias (see Sect. \ref{subsec:music_prof}), most of this difference is explained by the fact that our deprojection procedure cannot properly reproduce the spatial distribution of tSZ signal of disturbed clusters. Furthermore, clusters identified as relaxed might also be elongated along the line of sight. The deprojected pressure profiles of such systems are also slightly different from their true pressure profile extracted from the simulated cube. Using a triaxial deprojection method \citep[{e.g.,}][]{ser12}, based on the combination of the NIKA2 SZ data and the \xmm\ X-ray data in the NIKA2 SZ large program, would be a way to include this systematic effect in the analysis.\\

The results shown in this section highlight the importance of an accurate measurement of the morphological properties of galaxy clusters at high redshift to probe a potential evolution of the intrinsic scatter associated with the mean pressure profile of a representative cluster sample at $0.5 < z < 0.9$ compared to the values measured at lower redshifts. As has been shown in \cite{rup18}, the instrumental performance of the NIKA2 camera facilitate the characterization of the morphology and ICM thermodynamic properties with high precision for radii enclosed between $0.1\mathrm{R_{500}} < r < \mathrm{R_{500}}$ for the redshift range considered in the NIKA2 SZ large program. We will therefore be able to study the impact of ICM disturbances, detected by the NIKA2 camera, on the mean pressure profile as obtained under the hypothesis of hydrostatic equilibrium. Furthermore, the use of multiwavelength data sets that will complement the NIKA2 SZ large program will allow us to develop new analysis tools to take into account the systematic effects highlighted in this paper on the measured distribution of normalized pressure profiles.

\section{Conclusions}\label{sec:Conclusions}

We have used synthetic clusters from the MUSIC simulation to build a twin sample of the one that is currently considered for the NIKA2 SZ large program. The clusters were selected in order to avoid biasing the sample toward a given morphological state because the fraction of morphologically disturbed clusters is still poorly known at $z>0.5$. For each selected cluster, we computed the biases on the mass and integrated Compton parameter based on the true mass profile and that estimated under the assumption of hydrostatic equilibrium using the gas density and pressure profiles extracted from the simulation. We show that the distributions of these biases share the same median values for the subsamples of morphologically relaxed and disturbed clusters. However, the scatter associated with these distributions is significantly different for the two subsamples. If we consider the subsample of disturbed clusters, the fact that we rely on the hydrostatic equilibrium assumption to compute the integrated Compton parameter induces an additional scatter on this observable that is seven times larger than that obtained if the sample is only populated with morphologically relaxed clusters. While the actual value of this increase in the scatter strongly depends on the dynamical state of the selected clusters, this highlights the need to measure the hydrostatic bias parameter of each cluster in the sample to minimize this systematic effect that may significantly increase the intrinsic scatter associated with the $\rm{Y_{500}}${-}$\rm{M_{500}}$ scaling relation.

We simulated realistic NIKA2 and \planck\ tSZ observations of the selected synthetic clusters. The mock observations were analyzed with the NIKA2 tSZ pipeline to estimate the pressure profile of each selected cluster under the hypothesis of hydrostatic equilibrium. The deprojected profiles were used to estimate the mean normalized pressure profiles associated with the subsamples of relaxed and disturbed selected clusters.

We show that the NIKA2 tSZ analysis pipeline enables the recovery of the mean pressure profile of the relaxed and disturbed clusters at the percent level, although it relies on the hypothesis that clusters are objects in hydrostatic equilibrium. Furthermore, even though the number of selected synthetic clusters is lower than the number considered for the NIKA2 SZ large program, we find that we are able to detect $>2\sigma$ differences between the mean pressure profiles associated with the relaxed and disturbed subsamples in most of the radial range that can be constrained by NIKA2. Given the measured scatters on the distributions of pressure profiles for the relaxed and disturbed subsamples, we show that this significance would be even higher if the actual fraction of disturbed clusters in the NIKA2 sample is between 50\% and 70\%. We note however that this result strongly depends on the values of the measured scatters that are tied to the dynamical state of the selected clusters, which may vary significantly in the respective subsamples of morphologically relaxed and disturbed clusters in the NIKA2 sample.

We further show that ICM substructures that can be detected by NIKA2 have a significant impact on the mean pressure profile associated with the subsample of disturbed clusters. In particular, the intrinsic scatter of the pressure profile distribution associated with the selected subsample of disturbed clusters is 65\% greater than that observed for the profile distribution of dynamically relaxed clusters at $\mathrm{R_{500}}$. Furthermore, we show that the measured scatter on the mean normalized pressure profile associated with the whole selected sample is on average twice higher than the measured scatter associated with the true mean profile of the sample computed from the data extracted directly from the simulation. In particular, at $\mathrm{R_{500}}$, this increase is caused at 60\% by the normalization of the deprojected profiles with integrated quantities estimated under the assumption of hydrostatic equilibrium. Most of this increase in the scatter is however due to an incorrect modeling of morphologically disturbed clusters, which induces significant discrepancies between the shapes of their deprojected pressure profile and their true radial distribution of thermal pressure. This further impacts the estimates of the normalization quantities $\mathrm{R_{500}}$ and $\mathrm{P_{500}}$. We note that the galaxy clusters observed in the NIKA2 SZ large program will have different morphological properties that will lead to slightly different results on the intrinsic scatter of the measured distributions of pressure profiles. The fact that the NIKA2 SZ large program data will be complemented by X-ray and optical data sets will allow us to develop new deprojection procedures in order to minimize the systematic effects identified in this study. This will enable us to improve the characterization of the properties of the distribution of pressure profiles at high redshift.

If the self-similar hypothesis of cluster formation is not verified at high redshift, it will be essential to consider the intrinsic properties of the distributions of normalized pressure profiles, estimated on cluster samples in restricted mass and redshift ranges to minimize the bias associated with the pressure profile evolution on the constraints of cosmological parameters. The NIKA2 SZ large program will allow us to study the impact of high redshift ICM disturbances on both the shape and scatter of the pressure profile distribution. The high angular resolution of NIKA2 will be used to estimate the fraction of disturbed clusters in a representative sample of galaxy clusters at redshifts $0.5<z<0.9$ and to study its evolution with redshift. This task will not be trivial because it is challenging to define reliable morphological indicators that clearly separate relaxed and disturbed cluster populations from tSZ maps only. It will be necessary to combine morphological indicators specific to the X-ray maps of \xmm\ and the tSZ maps of NIKA2 to improve the criteria currently used to characterize the morphology of galaxy clusters. These joint analyses will benefit from the different dependences of these observables on the thermodynamic properties of the ICM and their integral along the line of sight. The combination of NIKA2 and \xmm\ observations will also enable us to estimate a potential deviation from the self-similar cluster formation processes by comparing the products of the NIKA2 SZ large program with the tSZ results obtained by other low redshift studies \citep[{e.g.,}][]{pla13}.

\begin{acknowledgements}
We would like to thank the anonymous referee for helpful comments and suggestions. This work has been partially funded by the ANR under the contracts ANR-15-CE31-0017 and by funding from Sapienza Universit\`a di Roma - Progetti di Ricerca Medi 2017, prot. RM11715C81C4AD67. Support for this work was provided by NASA through SAO Award Number SV2-82023 issued by the Chandra X-Ray Observatory Center, which is operated by the Smithsonian Astrophysical Observatory for and on behalf of NASA under contract NAS8-03060. The MUSIC simulations were produced with the Marenostrum supercomputer at the Barcelona Supercomputing Centre thanks to computing time awarded by Red Espa\~nola de Supercomputaci\'on. GY acknowledges financial support by the MINECO/FEDER in Spain through grant AYA2015-63810-P. We acknowledge funding from the ENIGMASS French LabEx.
\end{acknowledgements}

\end{document}